\documentclass[journal]{IEEEtran}
\usepackage{amssymb}
\usepackage{amsmath}
\usepackage{wasysym}
\usepackage{paralist}
\usepackage{cite}

\ifCLASSINFOpdf
  \usepackage[pdftex]{graphicx}
  \usepackage[font=small]{caption}
  \usepackage{subcaption}
\else
\fi
\usepackage{algorithmic}
\usepackage[ruled,slide,nofillcomment]{algorithm2e}

\usepackage{array}
\usepackage{url}

\begin{document}
\SetEndCharOfAlgoLine{}
%
\title{Information-centric Multilayer Networking: improving
  performance through an ICN/WDM architecture}
%
%
%

\author{Mays~F.~AL-Naday~\IEEEmembership{Member,~IEEE},
		Nikolaos~Thomos~\IEEEmembership{Senior Member,~IEEE}, 
    Martin~J.~Reed
\thanks{This work has contributed to, and is supported by, the EC H2020 ICT project POINT (643990). The authors are with the Department
of Computer Science and Electronic Engineering, University of Essex, Colchester,
Essex, CO4 3SQ UK e-mail: (mfhaln, nthomos, mjreed@essex.ac.uk).}
}

\maketitle

\tolerance=1000

\begin{abstract}
Information-centric networking (ICN) facilitates content
identification in networks and offers parametric representation of content semantics.
This work, proposes an ICN/WDM network architecture that uses these
features to offer superior network utilization, in terms of
performance and power consumption. The architecture introduces an ICN
publish/subscribe communication approach to the wavelength layer;
whereby, content is aggregated according to its popularity rank into
wavelength-size groups that can be published and ``subscribed to" by
multiple nodes. Consequently, routing and wavelength assignment (RWA) algorithms benefit from
\emph{anycast} to identify multiple sources of
aggregate content and allow optimization of the source selection of
light-paths. A power-aware algorithm, Maximum Degree of connectivity
(MaxDeg), has been developed with the objective of exploiting this
flexibility to address the trade-off between power consumption and
network performance. The algorithm is also applicable to IP architectures, albeit with less flexibility. Evaluation results indicate the superiority of the proposed ICN architecture, even when utilizing conventional routing methods, compared to its IP counterpart. The results further highlight the performance improvement achieved by the proposed algorithm, compared to conventional RWA methods such as Shortest-path First Fit (SFF). 
\end{abstract}


\IEEEpeerreviewmaketitle

\section{Introduction}
Multilayer network architectures in the form of packet-over-wavelength
switched networks represent the core transport infrastructure of the
current, and the future, Internet. However, realizing optimized
resource management of such networking architectures has been
increasingly challenging; particularly with the shift of communication
patterns and dissemination strategies towards high frequency
information exchange, as well as the rapid growth of traffic volumes
associated with the high demand for video/audio content. This shift
introduces the need to utilize an information-centered communication
framework that has a better adaptability to the traffic needs.

Meanwhile, the growth of the Internet has been accompanied with a
dramatic change in aggregate traffic patterns. The most significant of
the changes is the convergence of the majority of the Internet
applications and services towards the use of the HTTP protocol;
whereby traffic is generated from Content Delivery Networks (CDNs),
clouds and HTTP caching points of large content/service providers
\cite{lab:trf, pop:ICnet,moc:trf}. A strongly correlated change in
usage behavior has also been witnessed in access networks
\cite{kih:trf}. This change provides motivation for developing a
scalable \emph{anycast} dissemination mode that offers flexible
resource utilization, realized for a broad range of traffic
granularities, and with minimal waste of resources. The change also provides the potential to introduce new aggregation mechanisms that are based on information semantics, such as content popularity, rather than end-point identifiers.

However, direct semantic-based information dissemination is not
natively supported within the current design of the Internet, due to
the host-centric nature of the IP model. Alternatively,
Information-Centric Networking (ICN) is an emerging approach that introduces the semantics of information to the communication process and thus facilitates the establishment of communication relations based on common information interest rather than end host identifiers \cite{tro:ICnet,jac:ICnetCACM,kop:ICnet}.
Consequently, it has the potential to bring new flexibility to the
resource management of multilayer networks through the ability to
manage traffic by the popularity of its content.

In particular, ICN facilitates traffic aggregation based on information semantics; this offers two major advantages. First, it allows for a native identification of multiple information sources, which can be used as cache points; thus, enabling a natural \emph{anycast} support in the network. This provides scalable and native information dissemination in a similar manner to CDNs without such architectural violations as those imposed by DNS and HTTP redirects. As a result, resource allocation may benefit from increased flexibility through inclusion of multiple sources. This, in turn, offers enhanced support for resilience, QoS and general traffic engineering functions.
Second, it allows for decomposing the demands in such a way that fits
the content requirements and the network constraints. These advantages, empower the network operator with enhanced management capabilities based on knowledge of the demand semantics as well as the network capacity.

Aggregating traffic based on information semantics requires the ability to represent the data properties and characteristics in organized structures.
In this context, ICN architectures such as PURSUIT
\cite{tro12:ICnet,tro:ICnet}, CCN~\cite{jac:ICnetCACM}
and DONA~\cite{kop:ICnet} offer flexible
aggregation of information at different based on different
semantics. In particular, the PURSUIT architecture allows any information semantic to be aggregated within multiple hierarchical, as well as multi-homed, data structures that can be published by any entity in the network (i.e. publisher, subscriber, network operator, etc). Therefore, it offers the network a direct method to describe traffic aggregation models based on information-semantics.

Although ICN architectures are well developed at the packet layer, any
future solution will require the use of
transport networks, such as optical switched networks, to provide the
underlying networking infrastructure. This is as yet, an unexplored
area within the ICN community. Consequently, the aim of this
paper is to use the information space and functional model of the PURSUIT architecture
to introduce, and control, a new multi-layer ICN/WDM network architecture. The proposed architecture utilizes these features of PURSUIT to: facilitate traffic aggregation based on information semantics; and, offer native support of \emph{anycast} delivery at the wavelength layer, resulting in enhanced light-path assignment approaches.

This paper considers the scenario of a large scale network operator
that has a number of alternative peering points and CDN caches. These
CDN caches may either be controlled by the operator or be placed in
the operator's network by a content provider. In this context, there
are often a number of alternative sources (ingress points) for large
scale content catalogs. Our work considers each set of sources with
the same content as represented using the same information identifiers
in an ICN context. The connection of these sources to points-of-presence in the network will be delivered through optical light-paths,
through the underlying transport network. By unifying the concepts of
traditional optical light-path planning and anycast, through ICN
naming, this paper demonstrates that there are significant
advantages over the contemporary IP/WDM architecture.

In particular it should be noted that here we are assuming that ICN is
used to categorize, and aggregate, content identifiers using an ICN
naming system. Consequently, the ICN architecture provides the
mechanism to bind content identifiers to a location, or multiple
locations. In contrast, contemporary IP carrier networks do not
provide this feature; therefore, content providers implement an
anycast like behaviour for clients through DNS/HTTP
redirection. However, when it comes to aggregate content
dissemination, there is no clearly defined mechanism in IP networks
for operators to identify aggregate content and provide traffic
management based on content type and location.

We elucidate our work using the PURSUIT architecture; however, the work proposed here is generally applicable to other ICN architectures with suitable modification. The most important features that enable this work are the ability to group information through naming and the ability to obtain data from, possibly, multiple locations (anycast), which are supported by most ICN architectures.
This brings us to the main problem addressed by this work, which is to
reduce the blocking probability and power consumption within this new ICN multilayer architecture
through routing and wavelength assignment that benefits from flexible source selection, provided by the new architecture. 

To frame the problem of source selection and RWA, we revisit the
emerging trade-off between network performance and operational
cost. This trade-off is illustrated by the high blocking rate and low
power consumption introduced by power-aware (PA) algorithms versus the
low blocking rate and, relatively, high power consumption, realized by least congestion path (LCP) algorithms \cite{wia:alloptic, coi:alloptic}.
We then show that by utilizing the ICN naming to aggregate content, superior performance can be realized for the same or lower power consumption, compared to that of the IP model.
The components of the novel ICN multilayer communication framework can now be introduced:\\
\textbf{ICN Multilayer Network Architecture} Using the PURSUIT architecture as a concrete example, we introduce the hierarchy of the ICN/WDM multilayer network; we present the architectural functions of each layer as well as their ICN pub/sub communication approach.\\
\textbf{Information-centric Network Model} Using a novel traffic aggregation model (based on information ``popularity'') that can be applied to the ICN/WDM architecture, we define a network model that describes the information-centric RWA (iRWA) problem within the ICN pub/sub communication approach.\\
\textbf{ICN Routing Algorithms} To solve the iRWA problem, we propose a novel power-aware algorithm that addresses the trade-off between power consumption and network performance. The optimization is carried out through joint use of asymmetrical, convex, cost functions that take into account the traffic aggregation model.

The rest of this paper is structured as follows: Section
\ref{sec:pursuit} outlines the PURSUIT architecture focussing on the key enablers for the ICN/WDM architecture. Section \ref{sec:ml} outlines conventional multilayer networking architectures and reviews state-of-the-art RWA algorithms addressing the trade-off between power-awareness and load. Section~\ref{sec:ml-icn}, introduces the ICN/WDM architecture and the pub/sub communication approach. Section \ref{sec:nm}, introduces the network model and defines the iRWA problem using popularity as one information semantic that can be utilized to aggregate content identifiers. Section \ref{sec:alg}, presents the algorithm for solving the iRWA problem, named Maximum Degree of connectivity (MaxDeg), with a summary of notations shown in Table~\ref{tab:not}; while Section \ref{sec:eva} evaluates the performance of MaxDeg within both the IP/WDM and the ICN/WDM architectures using Zipf's law as an exemplary popularity model to construct the ICN demand matrix. Finally, Section \ref{sec:con} draws our conclusions.

\section{Pursuing a Pub/Sub Internet (PURSUIT)}\label{sec:pursuit}
Here, we briefly outline the PURSUIT architecture~\cite{tro12:ICnet,tro:ICnet,xyl:ICnet} and highlight the features and functions utilized to support the proposed ICN/WDM architecture. 
The communication paradigm of PURSUIT is of a publish/subscribe
nature: content, service requests, message exchanges, etc. are information items that can either be published or subscribed to. This model is utilized for all communication, including those between the architectural functions. Information publications are arranged within hierarchies of groups named \emph{scopes} that reflect common semantics, such as: content type, security requirements, transmission constraints and so forth. Publications can be associated with each other and potentially be linked to any number of scopes, through which different semantics are reflected. A pub/sub relation is established when there exists a common interest of a particular publication, which would ultimately result in a data transmission from the information publisher to the information subscriber. 
The functional model of PURSUIT consists of three distinct core functions: \emph{Rendezvous}, \emph{Topology Management} and \emph{Forwarding}.\\
\textbf{Rendezvous (RV):}
maintains the information space and offers matching services between publications and subscriptions to the communicating entities. When a publisher wishes to provide information, it publishes its availability to the RV, which will position it in the information graph according to its scope affiliation and add the publisher to the set of publishers of this information. An interested subscriber wishing to receive this information sends a subscription request to the RV. When a match occurs, the RV publishes a topology formation request to the topology management function (TM) asking the latter to establish an end-to-end delivery path.\\
\textbf{Topology Management (TM):}
controls the network connectivity and handles topology formation tasks. Multilayer networks can be supported by establishing multiple TM functions, one per layer, that communicate with each other through the pub/sub model described above, as illustrated in \figurename~\ref{fig:mlt} and explained in Section~\ref{sec:ml-icn}. The TM selects the best publisher out of the set of publishers according to the optimization objective of the routing algorithm. When a delivery path is required, the LIPSIN mechanism~\cite{pet:ICnet} can be used; this requires the TM to create a Bloom Filter (BF) based Forwarding Identifier (FId), which will be provided to the publisher in order to specify the packets' route in the network.\\
\textbf{Forwarding (FW):} By default the PURSUIT architecture proposes
the LIPSIN mechanism~\cite{pet:ICnet} which is a form of source
routing. LIPSIN utilises
fixed length link identifiers (LIds), rather than using node
identifiers. The TM forms the FId by encoding each of the path LIds in
the BF and this FId is carried in the packet header. Consequently, each forwarding node in the network can
accomplish simple and fast packet processing by performing bit-wise AND and COMPARE operations to test the membership of the node's LIds in the received FId.

Various optimization efforts have been made within PURSUIT, including:
resource optimization through routing over multiple links based on information semantics, thereby maximizing the free
capacity~\cite{mjr:te}; source recovery by exploiting the advantages
of \emph{anycast} to realize information as well as network
resilience~\cite{nad:ICnet}; and, caching optimizations based on
information placement and routing \cite{sou2:zipf}.

\section{Conventional Multilayer Architectures}\label{sec:ml}
Core networks are generally decomposed into multi-layer systems. This
is a natural decomposition as physical connectivity is naturally
circuit oriented (e.g. a fiber optic cable), while packet transport is
generally switched on a per-packet basis either according to routing
tables or virtual circuit switch tables. This paper considers a two layer approach to resource management, where the packet switched layer is connected using an optical switched layer that can switch light-paths between optical cross-connects (OXCs). The OXC layer has strong integer constraints on the path computation due to the limited light-path resources and the inability to support multi-granularity bit-rate per light-path. In addition we assume that the wavelength of a single path cannot be changed in an OXC: a wavelength continuity constraint. In contrast, the packet switch layer allows almost infinitesimal control over how packets are switched and shared over the circuits that provide connectivity between end-points in the network. Thus the problem is to determine light-paths that provide the necessary transport pipes to the packet switched layer and that meet the required optimization goals. 

\subsection{Routing and Wavelength Assignment in IP/WDM networks}\label{rwa:ip}
The problem of light-path provisioning (i.e. RWA) has been extensively studied for IP/WDM architectures with the general objectives of conserving wavelength resources while maximizing the network performance~\cite{raj:alloptic,zan:alloptic,and:rwa,ram:rwa,ban:rwa}.
Typically, these objectives can be realized by utilizing load-balancing algorithms such as Shortest-Path First Fit (SFF)~\cite{dij:rwa} and Least Congested Path (LCP)~\cite{zan:alloptic,ozd:rwa,cha:rwa,chu:rwa}, which evenly distributes the load across the network links and therefore reduces the chances of developing bottlenecks.
Although load balancing significantly improves the network performance, it comes at the cost 
of poor utilization of power hungry components such as Erbium Doped Fiber Amplifiers (EDFA) and 3R in-line generators~\cite{coi:alloptic}.
In particular, LCP algorithms generally tend to ignite most of the free links in the network before efficiently utilizing active links. Therefore, in scenarios where the network is lightly loaded, a large number of components operate with a small amount of load per link. This leads to a number of undesired consequences such as: high power consumption, high operational costs and under utilization of operating elements, especially when the overall offered load is relatively low~\cite{coi:alloptic}. 

Power consumption of Telecoms infrastructure has been rapidly
increasing with the expansion in the underlying transport networks,
such that it is now key optimization challenge in current as well as future infrastructures~\cite{mud:pow,zha:pow}. 
~Power consumption can be reduced by utilizing Power-Aware (PA) algorithms, which increase the utilization of active links in the network in order to decrease the number of operating components~\cite{yon:rwa,coi2:rwa}.
These algorithms perform well when the offered load is low to medium. However, when the load increases, such algorithms tend to create bottlenecks by overloading active links while leaving others free or under utilized. This imbalance in the load distribution often results in a sharp degradation of the network performance, especially when the offered load to the network is relatively high. 

A trade-off between power saving and network performance, has been discussed in \cite{wia:alloptic, coi:alloptic} highlighting the advantage of adopting both aspects in suitable cost functions to mitigate the impact of the power optimization objective on the network performance. However, these algorithms generally result in blocking rates that are higher than SFF or LCP.
In this paper, we address this shortcoming by developing, in
Section~\ref{costfun}, a new algorithm that utilizes asymmetric
convex functions that offer better network performance than SFF for lower power consumption than LCP. 
We also illustrate the impact of the IP model limitations on the proposed algorithm, which causes the resultant power consumption to be always higher than that of SFF; while the added flexibility of ICN, illustrated in the next section, significantly enhances the outcome of the algorithm. This will be illustrated by a higher performance of the proposed algorithm compared to that of SFF for lower, or similar, power consumption. Furthermore, we will show the potential of an ICN architecture to enhance the performance of conventional RWA algorithms, such as SFF and LCP.

\section{An ICN Multilayer Architecture}\label{sec:ml-icn}
\begin{figure}[tb]
 \centering
 \includegraphics[width=\columnwidth]{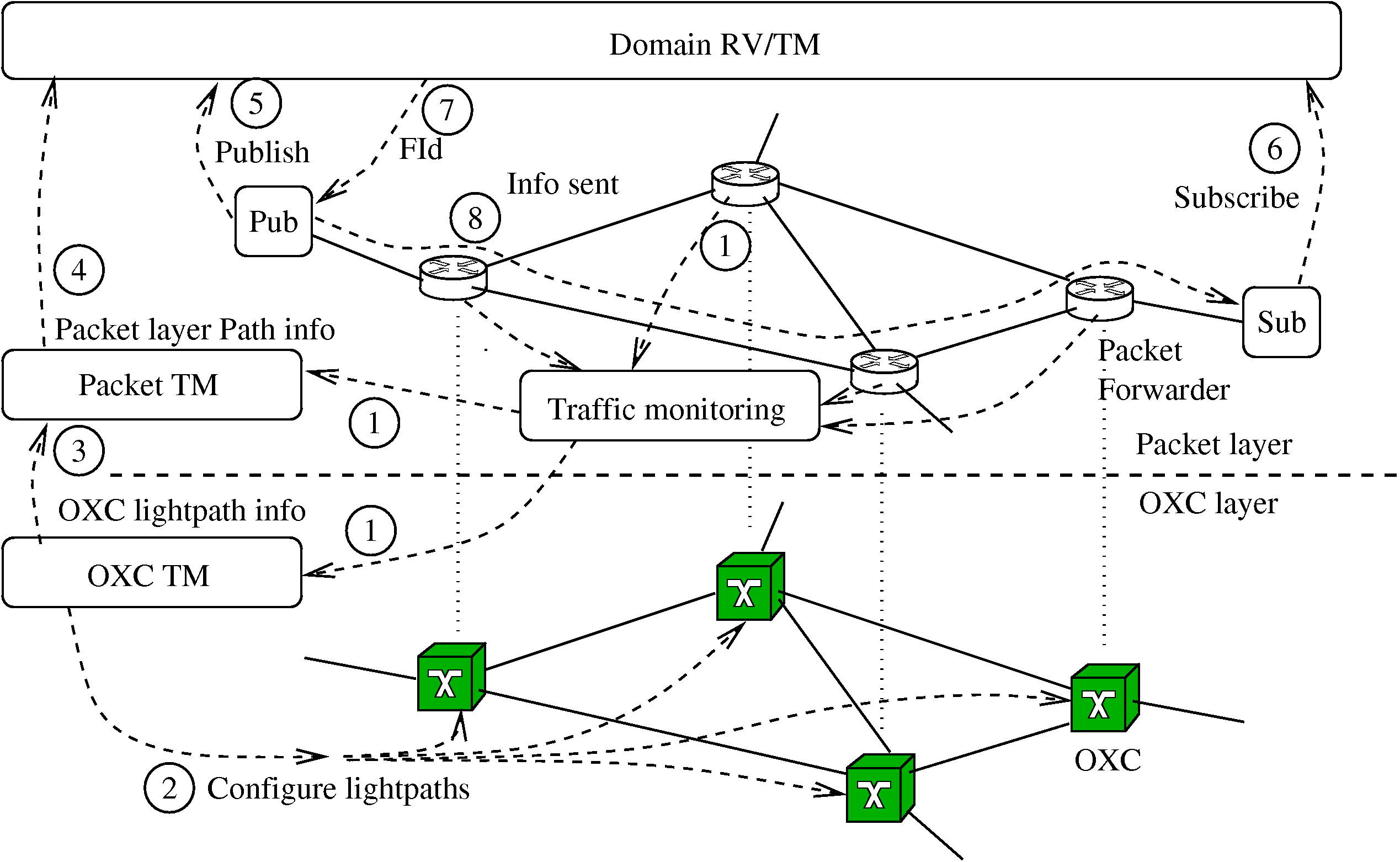}
 \caption{ Multilayer topology management showing a hierarchical relationship between TMs. The numbered arrows show information flow in a pub/sub relationship and the conceptual order of the information flows.}
 \label{fig:mlt}
\end{figure}

The architecture of the proposed solution is shown in \figurename~\ref{fig:mlt}. This shows how a physical node comprises both an OXC and packet switched component, but at separate layers. This is a rather simplistic view as there may be OXC systems without associated packet switching; or alternatively multiple packet switches may form a network of packet switches over a fixed optical transport network. Nevertheless, the result of this layering is that it leads to a hierarchical approach to topology management. In particular, \figurename~\ref{fig:mlt} shows part of the hierarchy in the form of a domain level TM which uses information from the packet layer TM, this in turn uses information from the OXC layer TM. Only one packet layer TM and OXC layer TM is illustrated in the diagram, but it might be that multiple areas of the network, possibly using different technology or forming different administrative zones, publish information that is used by the domain TM. 
The solution presented in \figurename~\ref{fig:mlt} expands an ICN architecture to include the optical transport layer so that it is suitable for a carrier grade network. The numbered arrows on the diagram show the conceptual sequence of events. Forwarders monitor traffic at their ingress nodes and publish this information as Step 1. While in theory it would be possible to report individual forwarder queue occupancy for Traffic Engineering (TE) purposes, for a complex network, buffer occupancy generally changes too quickly for a centralized system to respond to this change and maintain a stable control system. 
Consequently, the monitoring aims at measuring general traffic trends over fixed time periods, e.g. in the range of few minutes.
This periodic traffic information is published and the packet and OXC layer TMs subscribe to this information. The OXC layer is typically not very agile as the time to change a path may take the order of tenths of a second which disrupts existing traffic. Consequently, changes are made infrequently (in order of hours or days) to avoid too much disruption. The OXC layer uses the traffic monitoring information and network planning information to make path computation decisions in the OXC-TM and setup the light-paths that provide the connectivity to the packet layer. The slow changing information on the optical light-paths is then published and the packet layer TM subscribes to this information in Step 3.

The packet layer TM computes optimum Forwarding Identifiers (FIds)
between each pair of nodes in the network based upon the predicted
traffic from the traffic monitoring information that it has
``subscribed to". These FIds are calculated in advance of when they
will be used through the optimization algorithms that meet the
required TE goals; these algorithms are described later. In carrier networks, there is evidence that aggregated traffic is stable over the course of minutes \cite{Zha:te}, so that this FId calculation can be made every few minutes and still maintain a solution that is close to optimal. This FId information is published in Step 4 and the domain TM subscribes to this information and caches the FIds for each node pair. In the case of \emph{multicast} traffic these FIds can be combined to form trees. The result of this cooperation between the three TMs is that the domain TM can make a very fast FId calculation based upon a simple lookup table and yet supply paths that meet the designated TE goals. In the diagram, the publisher/subscriber nodes are notionally connected to the ingress/egress of this network. However, in practice this is unlikely as, other networks, with their own topology management function, may be between the publisher/subscriber and the ingress/egress. Here we investigate the intra-domain scenario as shown in \figurename~\ref{fig:mlt} assuming that the inter-domain problem is essentially a collection of intra-domain optimizations that are considered independently by each operator, as with the current Internet.

The path computation in the packet layer, highlighted in Section~\ref{sec:pursuit}, has an important advantage with respect to TE that will now be described. In IP or IP/MPLS networks, traffic is generally transmitted over a single path due to the nature of TCP, which generally performs poorly if traffic belonging to the same micro-flow is transported over different links. This is fundamentally linked to the fact that IP uses a destination based identifier for its forwarding model. In IP, it is possible to switch on micro-flows, but these can only be discriminated by utilizing network/transport layer inspection and maintaining some per-flow state. In core networks, this is prohibitively expensive. In contrast, the functional model of the PURSUIT architecture proposes a different forwarding model where the FId uniquely identifies the path rather than a destination.
This is made possible using the source routing LIPSIN approach
described earlier~\cite{pet:ICnet} that allows each publication to be
sent over a uniquely defined path. Consequently the packet layer TM
can report multiple paths between each pair of nodes when it publishes
the FId information (that the domain TM subscribes to). In addition,
it publishes how the FIds should be used; for example how much traffic
should be allocated to each FId to meet the TE goals. This association
between a path and a publication, combined with the ability to
aggregate publications based on their popularity, allows for
introducing popularity semantics to the wavelength resources; thereby,
supporting native anycast dissemination directly on the optical layer, as will be described in the next section.

\section{From a Host to Information Centric Network}\label{sec:nm}
Here, we present a migration story of the demand formulation from the host-centric model of the IP/WDM architecture to the ICN model. The new formulation defines ``popularity" as a suitable information semantic for the purpose of traffic aggregation, which allows straightforward support of \emph{anycast} delivery.
\begin{table}[tb]
	\centering
	\caption{Summary of Notation and Definitions}
	\label{tab:not}
	\begin{tabular}[\columnwidth]{|p{0.17\columnwidth}|p{0.73\columnwidth}|}
		\hline
		Notation & Definition\\
		\hline 
		$G(V,E,W)$ & Graph: vertices $V$, edges $E$ with set 
                             $W$ wavelengths\\ 
		\hline
		$e_{v_a \rightarrow v_b}$ & Edge $e \in E$ connecting vertex $v_a$ to $v_b$\\	
		\hline
		$B,U$ & The wavelength and optical channel capacities of $G$\\
		\hline
		$X,\ Z, \ M$ & Constants: $X = |V|$, $Z$ the number of
                               wavelengths in a node $v_a \in V$, $M$ the number of light-publications\\
		\hline
		$D, \ D', \Delta$ & The conventional IP Demand set, the transformation of the IP demand to include information identities, and the ICN demand set\\
		\hline
		$\Gamma, \theta_e$ & $\Gamma$ is the set of flows in conventional IP/WDM architectures. $\theta_e$ is the capacity occupation on an edge $e$ \\
		\hline
		$F(e), \chi(p,i)$ & $F(e)$ is the cost function of edge $e$. $\chi(p,i)$ is the cost of selecting $p$ to provide $i$.\\
		\hline
		$ \zeta_p^i$ & The relay-distance cost of selecting $p$ to provide $i$.\\
		\hline
		$I', \ I, \ k$ & $I'$ the set of light-publications of
                                 $D'$, $I$ the global set of light-publications, $k$ is the rank of an $i \in I$\\
		\hline
		$P,\ S$ & Global set of light-publishers $P$ and
                          light-subscribers $S$\\
		\hline
		$\Lambda$ & The global set of established pub/sub relations\\
		\hline
		$\alpha^i_p, \ \beta_p$ & $\alpha^i_p$ the fraction of relations provided by $p$ that corresponds to $i$, $\beta_p$ the ratio between the number of established relations from $p$ to the number of published light-publications by $p$\\
		\hline
		$L(k,M,\epsilon), \epsilon$ & Zipf probability mass
                                              function 
                                              with exponent $\epsilon$\\
		\hline
	\end{tabular}
\end{table}
\subsection{Network Definition and Problem Statement: The Host-Centric Perspective}
A wavelength routed network is modeled as a directed graph $G(V,E,W)$
that consists of a set of vertices $V= \{v_1,v_2,\dots\}, \ |V|=X$,
connected by a set of edges $E= \{e_{v_a \rightarrow v_b}\mid v_a,v_b
\in V, v_a \neq v_b \}$; whereby, each edge $e$ is defined by its
source-sink pair of vertices $v_a,v_b$. Each $v_a \in V$ decomposes
into: a packet forwarder (FW) and Optical Cross-connect (OXC). For
the sake of simplicity, we assume each FW maintains the same set of
wavelengths $W= \{w_1,\dots,w_\kappa,\ldots\}, \ |W|=Z$, and that these
wavelengths connect the FW to its corresponding OXC which facilitates wavelength switching through the network. In this model, the OXC is an optical channel provider that facilitates wavelength connectivity between adjacent FWs.

The node architecture constrains the packet layer forwarding by the
number of transceivers deployed in each FW, which is assumed to be
fixed $\forall v_a \in V$. We define the total wavelength
capacity, $B$, of the graph, $G(V,E,W)$, as the total set of wavelengths provided by each $v_a \in V$:
\begin{eqnarray}
B & = & \{B_{v_a,w_\kappa} \mid v_a \in V, \ w_\kappa \in W\}\label{c11}
\\
|B_{v_a}| & = &  Z
\\
|B| & = & \sum_{v_a \in V} |B_{v_a}| = Z X\label{c21}
\end{eqnarray}
where $B_{v_a,w_\kappa}$ denotes the presence of wavelength $w_\kappa$
on node $v_a$. The wavelengths available in each node is represented
by the set $B_{v_a}$ and the number of wavelengths is denoted by $Z$.
Equation~(\ref{c21}) constrains the wavelength capacity, $B$, of the
graph to the total number of wavelength transceivers installed in the
network of $X$ nodes, which enforces an upper bound on the total number of light-paths that can be provisioned through the network. 

The number of optical channels provided by each OXC is constrained by the minimum of: 
\begin{inparaenum}[\itshape a\upshape)]
\item the number of wavelengths supplied by the FW of each $v_a \in V$; and,
\item the number of input-output pairs of ports provided by the OXC of each $v_a \in V$.
\end{inparaenum}
When the OXC is a non-blocking switch, the number of optical channels provided on each edge $e$ is only constrained by the number of wavelengths, $Z$, supported by the connected FWs.
Thus, we define the optical channel capacity, $U$, as the total set of optical channels provided on each edge $e$ connecting $v_a$ to $v_b$:
\begin{eqnarray}
U &=& \{U_{e_{v_a \rightarrow v_b},w_\kappa} \mid e_{v_a \rightarrow v_b} \in E, \ w_\kappa \in W \} \label{c1}
\\
|U_{e_{v_a \rightarrow v_b}}| & =& Z \label{ce1}
\\
|U| & = &  \sum_{e_{v_a \rightarrow v_b} \in E} |U_{e_{v_a \rightarrow v_b}}| = Z |E| \label{c2}
\end{eqnarray}
where $U_{e_{v_a \rightarrow v_b}, w_\kappa}$ represents the availability
of an optical channel that can carry wavelength $w_\kappa$ on edge $e_{v_a \rightarrow v_b}$; and, $U_{e_{v_a \rightarrow v_b}}$ is the
total optical channel capacity of the edge $e_{v_a \rightarrow v_b}$. For simplicity, in the rest of the paper we will refer to $w_\kappa$ as $w$ and $e_{v_a \rightarrow v_b}$ as $e$.

Equation~(\ref{ce1}) defines each edge in the network as a
single fiber link that has a total number of optical channels equal to the number of
wavelengths, $Z$, provided by the connected vertices. When multiple fiber links connect a
pair of OXCs, it is translated into multiple edges between a pair of vertices.
Accordingly,~(\ref{c2}) constrains the optical channel capacity, $U$, of the graph to the sum of the optical channels provided on every edge. This quantity defines the switching capacity of the network, which reflects the flexibility of moving wavelengths across the graph (i.e. wavelength reuse). 
Noticeably, when the OXC is a blocking switch, $U$ further constrains the number of light-paths that can be provisioned in the network.

Next, we define the offered load to the network as a set of demands (commodities)
\textit{D}, where each $d \in D$ is a tuple of source $p$, destination $s$ and value $n$: 
\begin{eqnarray}
D & = & \{\langle p,s,n\rangle \ \mid p,s \in V, p \neq s , 1 \leq n \leq Z \}
\\
 \sum_{d \in D} n & \leq & \min( |B|, |U|) \nonumber
\end{eqnarray} 
where $n$ is an integer value that represents the number of
requested light-paths from $p$ to $s$, which is bounded by $Z$, the total number of
wavelengths maintained by each vertex. The formulation of the constraint stems from the fact that when a light-path is provisioned from $v_a$ to $v_b$, an optical transmitting port is reserved on $v_a$ while an optical receiver port is reserved on $v_b$. This implies that the ports in the opposite direction (i.e. the transmitting port of $v_b$ and the receiving port of $v_a$) are free to satisfy other requests. Accordingly, the total number of requested light-paths by $D$ is bounded by either the wavelength or optical channel capacity, depending on whether the OXC is a blocking or non-blocking switch.

$D$ can be satisfied by establishing a set of light-paths denoted as flows $\Gamma$ that will be defined below for each light-path request in $d \in D$ on a wavelength $w \in W$ of an edge $e \in E$:
 \begin{eqnarray}
\Gamma &=& \{\gamma_{e,w}^d \mid d \in D, e \in E, w \in W \}
\\
\gamma_{e,w}^d & =& \left\{
 \begin{array}{l l}
  1, & \textrm{if } U_{e,w} \textrm{ is reserved for } d\\ \label{eq:ipflow}
  0, & \textrm{otherwise}
 \end{array} \right.
 \end{eqnarray}
 Therefore, the capacity occupation on any edge $e$ by the collection
 of flows passing through it, is:
 \begin{equation}
 \theta_e = \sum_{w \in W} \sum_{d \in D} \gamma_{e,w} ^d
 \end{equation}
We now define the cost of assigning a new flow to an edge $e \in E$ as a function $F(e)$ of the capacity occupation on $e$. Now the problem of RWA can be stated as:

\emph{Objective:}
 \begin{equation}
 \min \sum_{e \in E} F(e)\label{eq:F_e}
 \end{equation}
\emph{ Subject to:}
 \begin{inparaenum} 
 \item the wavelength continuity constraint, 
 \item the optical channel capacity constraint $ \theta_e \leq |U_e|$, and,
 \item the flow conservation constraints.
 \end{inparaenum} 
 
The cost function, $F(e)$, can be selected to achieve the optimization of a desired aspect. 
For example, if one wishes to minimize the number of hop-counts only
(as in SFF algorithm), then it is appropriate to set $F(e) = 1 $ when
$ \theta_e < |U_e| $. 
This work proposes a power-aware $F(e)$, in Section~\ref{costfun},
that addresses the trade-off between congestion and power consumption,
whereby power expenditure is reflected by the ratio of inhabited fiber
links in the network. For each fiber link, power usage is mainly
derived by power-hungry pre- and post-transmission amplifiers as well
as 3R in-line regenerators. The focus is set on minimizing the number
of under-utilized fibers in order to reduce the overall number of
active amplifiers, without jeopardizing the network performance. The
proposed function is utilized to improve the resource utilization of
the IP/WDM architecture.

So far we have defined the standard model for conventional IP/WDM
networks. In the following sections we will adapt this nomenclature to
encompass the ICN demand model and formulate the problem of
an information-centric variant of RWA (iRWA). In particular, for the
ICN/WDM architecture, we formulate the ICN demand set and expand the
optimization scope to include optimized source, as well as path,
selection.

\subsection{Information Centralism}\label{sec:icndefs}
The previous definition of the demand matrix does not provide any semantics to the relation between the communicating nodes, apart from the number of light-paths requested between them. Hence, the network has no knowledge of the content carried in each aggregate flow and whether they share common semantics. Therefore, it cannot detect duplicate content transmissions or benefit from content visibility (popularity) to optimize the resource utilization.
This is a reasonable compromise for IP/WDM networks, considering that aggregating traffic based on TCP/UDP sessions, to emulate content recognition, is a cumbersome mechanism that is neither feasible nor scalable.

However, the proposed ICN/WDM architecture benefits from the hierarchical, semantic-based, identification of information, offered by ICN. This hierarchy allows for establishing different classifications of an information identifier by different publishers/subscribers, each of which reflects different semantics. Thereby, it lends the network to semantic-aware solutions that are based on customers' needs. 
Here, we exploit this hierarchy to aggregate individual pub/sub relations of raw publications such as video clips, emails and so forth on the basis of information popularity, rather than end host location.
In this aggregation model, an individual publication has an additional, popularity-based, scope affiliation that indicates the publication popularity based on certain criteria. Here, we define this criteria as the number of subscriptions and/or advertisements made for the publication. Publications that have a particular number of advertisements/subscriptions are grouped within a popularity scope of a specific rank. The ``popularity" scoping structure may consist of any number of ranking scopes, depending on the number of publications presented to the network. Each popularity scope aggregates a set of publications that is large enough to occupy the bandwidth of a single wavelength. As a result, in the ICN/WDM architecture, the value associated with each demand does not only reflect the number of requested light-paths, but also the associated set of popularity scopes.

In a multilayer carrier network, publications and subscriptions are made by core forwarders at the level of popularity scopes, which is equivalent in granularity to a request for a single light-path. Accordingly, we introduce the notion of: \emph{light-publication} to denote a popularity scope of specific rank, \emph{light-publisher} to refer to such a node that advertises a certain light-publication(s) and \emph{light-subscriber} to refer to such a node that subscribes to a particular light-publication(s). When a mutual interest occurs in a light-publication, a ``light-pub/sub" relation is to be established that involves the provision of a light-path from the light-publisher to the light subscriber.
This allows the IP demand matrix to be transformed into the corresponding ICN matrix, in which each element reflects a common interest of a particular light-publication.

 \subsection{The Information-centric Model and Formulation}\label{sec:icnmodel} 
To develop a structural transformation from the existing IP model
towards ICN, we start by reformulating the IP demand matrix to include
information identities that represent \emph{light-publications}. We
will represent this as a modified demand set $d' \in D'$. We will then develop
this further by noting that there may be multiple sources,
\emph{light-publishers}, and this requires a further modified demand
set that we will define as $\delta \in \Delta$.

\subsubsection{Developing the modified, single source demand matrix}
First we define the set of all information identities,
namely light-publications, as the set $I$. Consequently, we redefine the demands $d\in D$ as
$d'\in D'$, which denote a relationship between a set of
light-publications and publishing/subscribing nodes. Thus, $d'\in D'$
is represented as a tuple of $\langle p, s, n, I'\rangle$, where $I' =\{i_1, i_2,
\ldots\} \subseteq I, \ |I'|=n$ is the set of light-publications offered
by $p$ to $s$. 
Each light-publication $i \in I'$ aggregates a set of content publications that have a common popularity rank. To determine the capacity requirements of light-publication $i$, we propose summing the throughput requirements of all the content publications aggregated within $i$. This is a reasonable approach as there is mounting evidence that aggregate traffic in large carrier networks behaves as a Poisson process at the time frames used by network management systems, i.e. few minutes~\cite{kar:te,nuc:tm}. Accordingly, for any wavelength $w$ on an edge $e$ we constrain the capacity requirements of each $i \in I$ to be less than or equal to $U_{e,w}$.
This constraint ensures that the capacity required by $i$ does not exceed the offered capacity of a single wavelength. Accordingly, the satisfaction of a request for light-publication $i$ may have a similar binary representation to that of~(\ref{eq:ipflow}).

In the ICN architecture, a single light-publication $i$ may be offered
by multiple light-publishers and ``subscribed to" by a number of
light-subscribers. For example, a popular content provider may have a
number of points-of-presence to provide content. Thus, a light-publication becomes the
commodity upon which light-pub/sub relations are
established. Therefore, the ICN demand matrix $\Delta$ can
be developed from the formulation of $D'$ to reflect common interest
in each light-publication as will now be described.
As previously defined, $I = \{i_1, i_2, \ldots\}$, $|I|=M$, is the global
set of light-publications; items from $I$ are advertised in the
network by one or more publishers from the set of all light-publishers $P =
\{p_1, p_2, \dots, p_h\} \subseteq V$, and each publisher may advertise
a particular subset of light publications $ I_p = \{i_1, i_2,\dots, i_q\}, \ I_p \subseteq I \label{ipub} $. Each of these light-publications can be
subscribed to by one or more subscribers from the set of all
light-subscribers $S = \{s_1, s_2,\dots, s_o\} \subseteq V$.

At initialization, when there are no light-subscriptions
satisfied (i.e. no light-paths provisioned in the network), $S_0 \cap
P_0 = \emptyset$. This stems from the fact that it is illogical for a
publisher of an information to also subscribe to it. However, at a
later point in time when there are one or more light-subscriptions
that have been satisfied, the corresponding light-subscribers may act
as light-publishers, caches, of the provided light-publications to other
light-subscribers. Therefore, more generally, it becomes likely that
$S \cap P \neq \emptyset$. 
Notably, the definitions of $I$ and $I_p$ imply that there may exist two or more light-publishers, as in~\figurename~\ref{fig:icndm}, $p_a, p_b \in P$ that have published a common subset of light-publications. Similarly, there may exist two or more light-subscribers, \emph{i.e.} $s_a, s_b \in S$ that have subscribed to a common set of light-publications. 
 
\subsubsection{ICN demand matrix}
\label{sec:icn-demand-matrix}

Now, for each light-publication, $i$, we define: the common set of
light-publishers, $\rho_i \in P$ and the common set of light-subscribers,
$\sigma_i \in S$. Accordingly, we can define the ICN demand matrix as a set of demands $\Delta$; where each $\delta_i \in \Delta$ is a tuple $\langle \rho_i, \sigma_i, i \rangle, \ \forall i \in I$ such that: 
\begin{eqnarray}
\Delta & = & \{\langle \rho_i, \sigma_i, i \rangle \mid \rho_i \subseteq P, \sigma_i \subseteq S, \forall i \in I \}
\\
|\Delta| &=&  M\nonumber
\end{eqnarray}
It should be noted that: since publication and subscription events of
a particular light-publication are
likely to be temporally decoupled, it is only when both events have
been registered by the RV that the demand is offered to the network by
the TM. Without loss of generality, we will assume that publications always
occur first; therefore, it is the subscriptions that cause the load to be offered to the
network. Accordingly, the number of subscriptions is constrained by
either the wavelength or optical channel capacity as formulated by:
\begin{equation*}
\sum_{\delta_i \in \Delta}|\sigma_i| \leq \min(|B|, |U|)\label{eq:sigma_con}
\end{equation*}

It is worth noting that $D'$, a single publisher/subscriber scenario, is a special case of $\Delta$, whereby: $|\rho_i| = 1, |\sigma_i|=1 , \ \forall \delta_i \in \Delta$.

\subsubsection{Problem Statement}
Now, the problem of information-centric routing and wavelength
assignment (iRWA) can be defined as determining the set of
light-publishers, routes and wavelengths that satisfies $\Delta$,
using the minimum amount of resources. It is known that the RWA
problem is, at best, NP-complete and hence it does not admit a fast
solution \cite{raj:alloptic,ste:rwa}. This is also true for the iRWA
problem that we examine in this paper. Thus, we decompose the iRWA
into two sub-problems: the problem of selecting a light-publisher to
form a light-pub/sub relation; and, the problem of finding a route
and a wavelength (i.e RWA) to deliver the content of the established relation. Although the problems are described
sequentially, they are solved jointly in the algorithm described in Section
\ref{sec:alg}. The two problems can be formalized as \emph{integer
 multicommodity flow} problems~\cite{ahu:netflow,raj:alloptic,
 ozd:rwa}. However, we introduce the notion of \emph{integer
 multicommodity relation} to denote the pub/sub nature of the
problem, which is different from conventional flow in IP.

Prior to formulating the problem of light-publisher selection, let us first define the set of established relations $\Lambda$ as:
 \begin{eqnarray}
\Lambda & = & \{\lambda_{p,s}^i \mid i \in I, p \in \rho_i, s \in \sigma_i\}
\\
\lambda_{p,s}^i & = & \left\{
 \begin{array}{l l}
  1, & \textrm{if } p \textrm{ satisfies } s \textrm{ demand for } i\\
  0, & \textrm{otherwise}
 \end{array} \right. \nonumber
 \end{eqnarray} 
Note that each $i \in I$ represents an identifier of the corresponding $\delta_i \in \Delta$, as $\delta_i$ denotes the matching publications and subscriptions of $i$.
Now, considering that $E_p \subset E$ is the set of outgoing edges of $p$, we define the number of relations of $i$, $\mu^i_{p,e}$, that can be established from $p$ over $e \in E_p$ as:
\begin{equation}
\mu^i_{p,e} = \sum_{s \in \sigma_i} \lambda^i_{p,s}, \ \forall p \in \rho_i, \rho_i \subseteq P
\end{equation}
 Now, we define the wavelength occupation of a light-publisher $p$ as:
 \begin{eqnarray}
\mu_p^i & = & \sum_{e \in E{_p}} \mu^i_{p,e}, \ \forall p \in \rho_i, \rho_i \subseteq P
\\
\mu_p & = & \sum_{i \in I_p}\sum_{e \in E_p} \lambda_{p,e}^i, \ \forall p \in P
 \end{eqnarray}
where, $\mu_p^i$ represents the wavelength occupation by the satisfied relations of $i$ from light-publisher $p$, and $\mu_p$ stands for the wavelength occupation by the overall set of established relations from $p$. Recall that $\sigma_i$ is the set of light-subscribers of $i$, and $I_p \subseteq I$ is the set of light-publications advertised by $p$.
 Next, we define the cost of selecting light-publisher $p$ to satisfy a new interest of $i$ as a function $\chi(p, i)$, described later in Section~\ref{costfun}. Now, we define the problem of selecting a light-publisher (i.e. the first part of the iRWA problem) as a constrained minimization problem, stated as:

 \emph{Objective:}
 \begin{equation}
\min \sum_{i \in I}\sum_{p \in \rho_i} \chi(p, i)
 \end{equation}
 \emph{Subject to:} Wavelength availability at a light-publisher of $i$:
\begin{eqnarray}
\mu_p \leq Z, \  \forall p \in \rho_i
\end{eqnarray}

In addition to selecting candidate light-publishers for a new
relation, there is the problem of routing the light-path that serves the pub/sub relation. Each $\lambda_{p,s}^i \in \Lambda$ can either be satisfied by: a single light-path, if there is a free wavelength across all the links from the light-publisher to the light-subscriber; or, by a concatenated sequence of light-paths of different wavelengths, whereby an intermediate FW facilitates the switching from one light-path to the next. Accordingly, the set of light-pub/sub relations of $i$, $\lambda^i$, can be recognized in the network by their associated light-publication $i$ as follows:
\begin{eqnarray}
\lambda^i & = & \{\lambda_{e,w}^i \mid e \in E, w \in W \}
\\
\lambda_{e,w} ^i & = & \left\{
 \begin{array}{l l}
  1, & \textrm{if } U_{e,w} \textrm{ is occupied by } i\\
  0, & \textrm{otherwise}
 \end{array} \right.
\end{eqnarray}
Now, the total capacity occupation on any edge $e$ can be redefined as:
\begin{equation}
\theta_e = \sum_{i \in I}\sum_{w \in W }\lambda_{e,w} ^i
\end{equation}
Thus, the cost of assigning a relation of $i$, $\lambda^i_{p,s}$, to a wavelength $w$ on an edge $e$ can be defined as a constrained minimization problem. Accordingly, the RWA part of the iRWA problem can be defined as:\\
\emph{Objective:}
\begin{equation}
\min \sum_{e \in E} F(e)
\end{equation}
\emph{Subject to:}
\begin{inparaenum}
	\item the wavelength continuity constraint;
	\item the optical channel capacity constraint, $\theta_e \leq
          U_{e,w}$ ;
	\item the relation conservation constraints of
          light-publication $i$, occupying an edge $e$ (recall that
          $e$ connects $v_a$ to $v_b$), which consists of two
          parts, the light-publisher constraint
		\begin{eqnarray}
		\sum_{w \in W} \lambda_{e,w}^i & =& \left\{
		\begin{array}{l l}
		\mu^i_{v_a,e}, & \textrm{if } v_a \in \rho_i\label{eq:icnflow}\\
		0, & \textrm{otherwise}
		\end{array} \right.
		\end{eqnarray}
		and, the light-subscriber constraint
		\begin{eqnarray}
		\sum_{w \in W} \lambda_{e,w}^i & =& \left\{
		\begin{array}{l l}
		1, & \textrm{if } v_b \in \sigma_i\\
		0, & \textrm{otherwise}
		\end{array} \right.
		\end{eqnarray}
\end{inparaenum}

\begin{figure}[tb]
  \centering
 \includegraphics[width=\columnwidth,keepaspectratio]{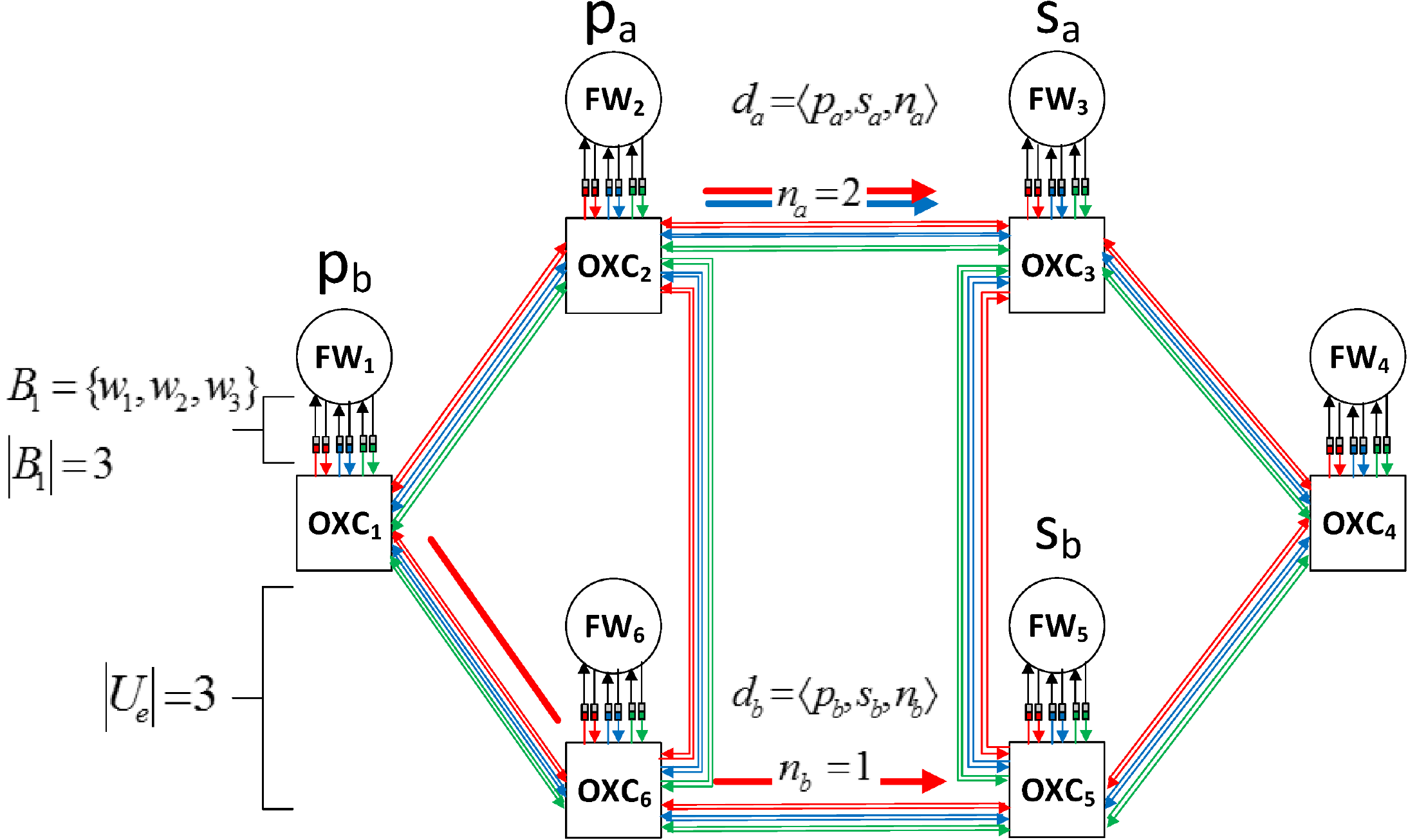}
  \caption{Demand requests in an IP/WDM network showing two connection requests between two pairs of nodes \{($p_a$, $s_a$), ($p_b$, $s_b$)\}.}
  \label{fig:ipdm}
 \end{figure}
\begin{figure}[tb]
  \centering
  \includegraphics[width=\columnwidth, keepaspectratio]{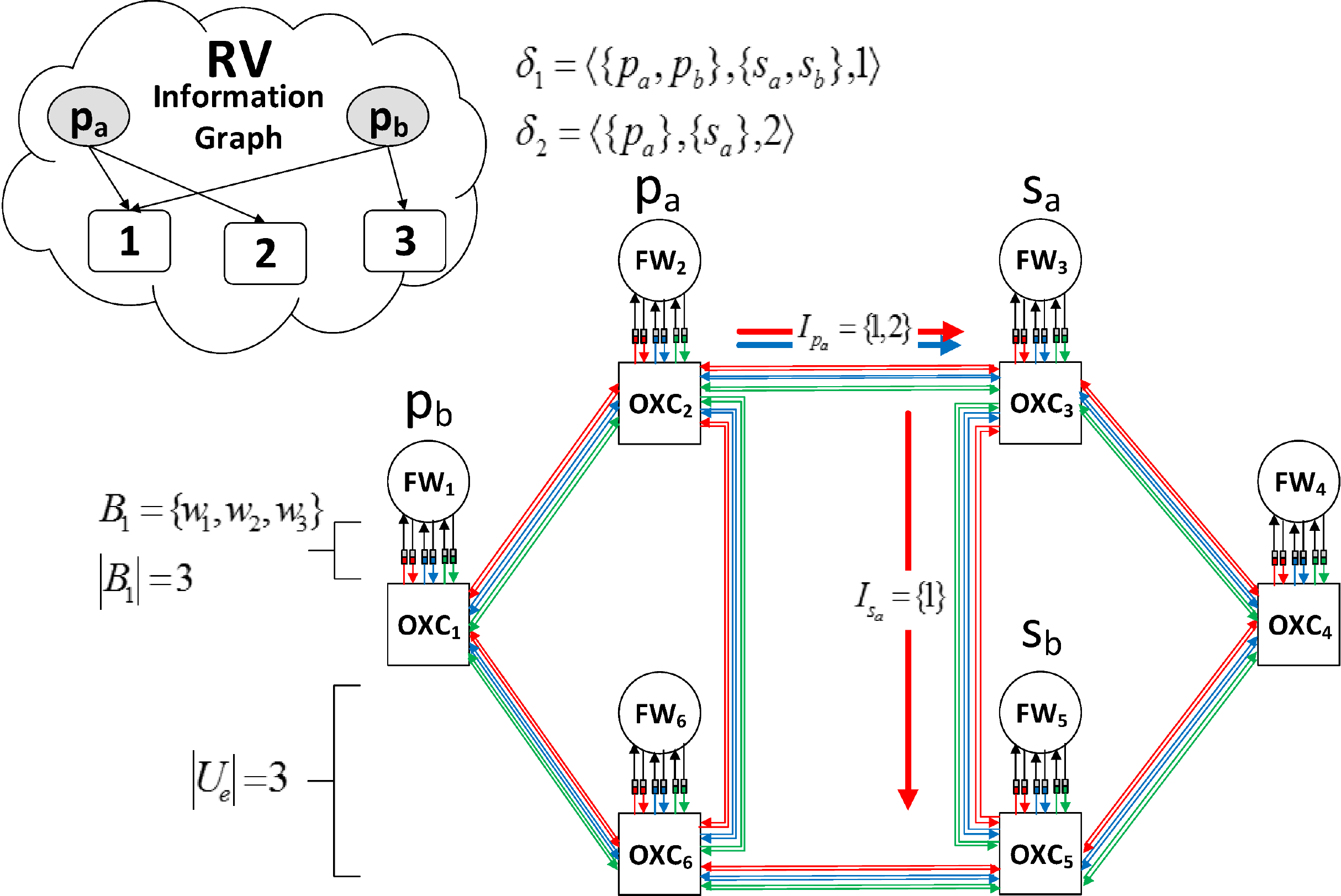}
  \caption{Pub/sub demand set in an ICN/WDM network showing two matching light-publications between two publishers \{$p_a$, $p_b$\} and two subscribers \{$s_a$, $s_b$\}.}
  \label{fig:icndm}
 \end{figure}

An example of an ICN demand set is illustrated in
\figurename~\ref{fig:icndm}, which is a transformation of the IP
demands presented in
\figurename~\ref{fig:ipdm}. \figurename~\ref{fig:icndm} shows that
$p_a$ and $p_b$ may advertise two sets of light-publications that
include a common element, \emph{light-publication $1$}. Both
light-subscribers $s_a$ and $s_b$ subscribe to light-publication
$1$. After a match has been triggered by the RV, the TM first selects
$p_a$ as the best candidate to provide light-publication $1$ to $s_a$
and accordingly provisions a delivery light-path from $p_a$ to
$s_a$. However, to provide light-publication $1$ to $s_b$, the TM has
a number of possibilities: for instance it could provide $1$ from
$p_b$, similar to the host-centric model. Although, this choice would
result in utilizing two data and light resources to transmit the same
information. A second option would be to utilize $s_a$ as a
caching/replication point that may act as a second light-publisher of
$1$; thereby, giving the RWA algorithm the option to select the best
light-publisher according to other criteria such as, for example, the
load (exhaustion) of each light-publisher. This choice would still
result in the utilization of two data and light resources, but it
comes with the advantages of optimized source selection for reduced
hop-count and source exhaustion. A third choice,
\emph{relay-forwarding}, would also utilize $s_a$ as a second
light-publisher; whereby, a light-path can be provisioned from $s_a$
to $s_b$. However, in this scenario $s_a$ would not cache the content
of light-publication $1$ and merely act as a \emph{relay-forwarder} or a patching
point between the incoming light-path from $p_a$ and the outgoing one
to $s_b$. This option represents a form of \emph{multicast}
dissemination; whereby, a data source and two light
resources are utilized to provide the same information. In this work
we have assumed that optical \emph{multicast} is not available,
however, if optical multicast~\cite{cha:omc} was available it could be
used; thereby, reducing the load on packet-layer forwarding. 
 
 The above options are design choices that depend, among other
 elements, on: the offered dissemination modes (i.e. \emph{unicast},
 \emph{multicast}), the forwarding mechanisms supported in the network, and
 the caching/replication policy/plan. Although, not all the options
 can be supported directly in any architecture, the second option in
 particular has distinct advantages in utilizing caching/replication
 features while maintaining compatibility with current forwarding
 mechanisms. In the PURSUIT ICN architecture, these choices have a
 direct support facilitated by the use of native \emph{multicast} forwarding
 mechanisms such as: LIPSIN \cite{pet:ICnet} and the multi-stage Bloom
 filter \cite{tap:bf}. Therefore, the solution proposed in this
 work supports both options two and three: namely a satisfied
 light-subscriber becomes a second light-publisher either by acting as a light-publisher in its own right, or through optical-multicast if it were to exist. 
It should be noted that the use of a second publisher is a form of
managed caching and we will see from the results in
Section~\ref{sec:eva} that this provides a significant
benefit. However, this work does not consider intelligent cache
management, which could bring even greater benefits.
 
\subsection{Power and Exhaustion Aware Cost Functions}\label{costfun}
Following the above formulations, two cost functions have been
developed: a link cost function and a light-publisher cost
function.
The functions described here were the result of extensive
experimentation; however, the authors do not claim their optimality.
The link cost function $F(e)$ has been defined as a convex asymmetrical function that minimizes the number of under-utilized links with the care not to jeopardize the network performance:
\begin{equation}
F(e) = \; \theta_e \mid 1 - \frac{Z^2}{ 4\theta_e(Z-\theta_e)} \mid / (Z-\theta_e)  \label{lcost}
\end{equation}
where $Z$ is the total number of wavelengths on any edge $e$. The first term of~(\ref{lcost}) is a parabolic function that handles the power aspect by creating a symmetrical cost variation with minimum cost reached when the capacity occupation on edge $e$ is $Z/2$, i.e., when the load on the edge is 50\% of its capacity. The second part of~(\ref{lcost}) is a positive ``exponential-like" function that achieves an improved load balance by assigning a monotonically increasing cost to each edge $e$, with increased capacity occupation $\theta_e$. Thus, the composite function is asymmetric. 

The light-publisher cost function attempts to select a light-publisher that reflects the best trade-off between the load introduced by a single light-publication and that of the total set of light-publications. Recall that the load induced by a light-publication $i$ is the result of the provisioned light-path(s) to establish a light-pub/sub relation(s) from a publisher of $i$ to the corresponding set of light-subscribers of $i$.
Consequently, we define two quantities: $\alpha_p^i$ and $\beta_p$. The quantity $\alpha_p^i$ is the fraction of established light-pub/sub relations, $\lambda_p^i$, from $p$ for light-publication $i$. This measure describes the exhaustion of a light-publisher by a single light-publication. Whereas, $\beta_p$ is the ratio of the overall established relations, $\lambda_p$, by light-publisher $p$ to the total number of advertised light-publications, by $p$. This quantity describes the general exhaustion level of a light-publisher. The two quantities are expressed as:
\begin{equation}
\alpha_p^i = \frac{\mu_p^i + 1}{\mu_p+ 1} ,\label{eq:alpha}
\quad
\beta_p  =  \frac{\mu_p}{|I_p|}
\end{equation}

The formulation of (\ref{eq:alpha}) prevents an undetermined value of $\alpha_p^i$ when the publisher does not have any established relations. Now, the cost of selecting a light-publisher $p$ to supply light-publication $i$ is defined as:
\begin{equation}\label{pcost}
\chi(p,i) = [e^{\alpha_p^i} -1]  e ^{\frac{-1}{|\beta_p -1|}}
\end{equation}
Note that for convenience in the later algorithm, $\chi(p, i)$ will be denoted as $\chi^i_p$ for a particular value of $p$ and $i$.

The first term ensures that the cost of a light-publisher increases when the number of established relations of a particular light-publication increases. Whereas, the second term ensures that the cost function follows a convex asymmetrical function with the increase in the ratio of established pub/sub relations to advertised light-publications, $\beta_p$. When the number of established relations is smaller than the number of advertised light-publications, the cost tends to decrease. However, when the number of established relations exceeds the number of advertised light-publications, the cost of using light-publisher $p$ starts to increase.

In the ICN architecture, when a satisfied subscriber of
light-publication $i$ acts as a relay-forwarder, a
\emph{relay-distance} cost is to be assigned to it; this determines
the ``goodness'' of the relay as a choice for provisioning the
required light-pub/sub relation of $i$. The relay-distance cost
describes how far the relay publisher is from the original publisher that provided the former with $i$. This is formulated as a ratio of the hop count between the relay and the original publisher, to the diameter of the graph $\diameter(G)$:
\begin{equation}
\zeta_p^i = hopcount(i,p)/\diameter(G)\label{rcost}
\end{equation}
 Notably, when comparing our proposed functions with existing PA cost functions, such as that of~\cite{coi:alloptic, wia:alloptic}, the performance and power saving of the latter's respective algorithms is comparable to that of SFF algorithm. It can be observed that although these algorithms offer considerable power saving compared to SFF, their performance is always lower than that of SFF. Whereas, the performance of our algorithm, described in the next section, is comparable to LCP and SFF algorithms; whereby, the network performance and power consumption of our algorithm is higher than that of SFF but lower than LCP, as will be illustrated by our evaluation analysis in Section~\ref{sec:eva}.

\section{Maximum Degree of Connectivity(MaxDeg): An Algorithm for Solving the iRWA Problem}\label{sec:alg}
Here we propose a greedy algorithm for solving the iRWA problem: Maximum Degree of connectivity (MaxDeg) shown in Algorithm~\ref{alg:icn_maxdeg}. The routing strategy of MaxDeg attempts to maximize the number of established relations for each light-publication.
 To achieve this goal, for each light-publication $i$ of $\delta_i \in \Delta$ (line \ref{alg:foralldelta}), MaxDeg attempts to maximize the number of light-subscribers $\sigma'_i \subseteq \sigma_i$ of $\delta_i$ that receives $i$. For each light-subscriber, $s \in \sigma_i$ (line \ref{alg:forallsigma}),
 the algorithm utilizes the cost functions of~(\ref{pcost}),
 and~(\ref{rcost}) to select the optimum light-publisher $p_{opt}$,
 while it uses the cost function of~(\ref{lcost}) in \emph{Dijkstra's} min-cost algorithm~\cite{dij:rwa} to select the optimum route, $\psi_{opt}$; depending on the location of $s$ and the current status of all light-publishers of $i$ (lines \ref{alg:forallrho}-\ref{alg:endforallrho}).
 Notably, when using the algorithm with IP, there is only one light-publisher.
 
 Once a light-publisher $p_{opt}$ and a route $\psi_{opt}$ are found (line \ref{alg:found}), the first available wavelength is assigned to the route by the first fit algorithm $FF(\psi_{opt}, U)$~\cite{raj:alloptic} (lines \ref{alg:ff}-\ref{alg:cache}).
 However, if there is no wavelength that is available on all the links from $p_{opt}$ to $s$, the route might still be provisioned using packet switching. 
 That is, if there exists a number of free wavelengths on parts of the route and sufficient wavelength capacity on intermediate FWs to switch from one wavelength to the next; thereby, allowing the route to be comprised of a patched sequence of light-paths. Note that only in the case of ICN, a subscriber can be added to the set of light-publishers as a relay (line \ref{alg:cache}). 
 On the other hand, if there are not enough free channel nor
 wavelength resources to establish a route from any $p \in \rho_i$ to $s$,
 the algorithm will not be able to select any $p_{opt}$; and
 therefore the light-pub/sub relation will be rejected (lines
 \ref{alg:block1} or \ref{alg:block2}). The number of accepted or
 rejected light-subscriptions are recorded using $A$ or $B$ respectively.
\begin{algorithm}[t]
\caption{$ICN: MaxDeg (G(V,E,W), \Delta, P, I)$}
\label{alg:icn_maxdeg}
\begin{algorithmic}[1]
\STATE $A \leftarrow 0; \ B \leftarrow 0; \ Z \leftarrow \mid W\mid;\
\theta \leftarrow \{0\mid \forall e \in E \}$
\STATE $U \leftarrow \{0 \mid \forall w \in W, \ e \in E\} ;\
F \leftarrow \{1 \mid \forall e \in E \}; \ $
\STATE $\mu \leftarrow \{0 \mid \forall p \in P, \ i \in I \}; \
\zeta \leftarrow \{0 \mid \forall p \in P, i \in I\};$
	\FORALL {$\delta_i \in \Delta$} \label{alg:foralldelta}
		\FORALL{$s \in \sigma_i$} \label{alg:forallsigma}
		\STATE $best \leftarrow \infty; \ good \leftarrow \infty; \ p_{opt} \leftarrow NA; \ \psi_{opt} \leftarrow \emptyset;$ 
		\FORALL[note for IP $|\rho_i|=1$]{$p \in \rho_i$} \label{alg:forallrho}
			\STATE$\alpha_p^i = (\mu_p^i + 1)/(\mu_p + 1); \ \beta_p = \mu_p / |I_p|$
			\STATE$\chi^i_p \leftarrow e ^{-1/|\beta_p -1|}[e^{\alpha_p^i} -1]$ \label{alg:pcost}
			\STATE$\zeta^i_p = hopcount(i,p)/\diameter(G)$\label{alg:dim}
			\STATE $\psi \leftarrow Dijkstra(G, p, s, F)$ \label{alg:dij}
			\STATE$good \leftarrow \chi^i_p [\zeta^i_p +\sum_{e \in
                          \psi}F(e)]$
			\IF{$good < best$}
				\STATE$best \leftarrow good; \ \psi_{opt} \leftarrow \psi; \ p_{opt} \leftarrow p$
			\ENDIF
		\ENDFOR\label{alg:endforallrho}
		\IF{$best \neq \infty$} \label{alg:found}
				\STATE $\eta \leftarrow FF(\psi_{opt}, U)$ \label{alg:ff}
				\IF{$\eta \neq \emptyset $}
					\FORALL {$[e,w] \in \eta$} \label{alg:pktst}
						\STATE $U_{e,w} \leftarrow 1; \ \theta_e \leftarrow \theta_e + 1;$\label{alg:update}
						\STATE $F(e)
                                                \leftarrow
                                                \theta_e \mid 1 - \frac{Z^2}{4\theta_e(Z-\theta_e)} \mid/(Z-\theta_e) $\label{alg:lcost}
						\STATE
                                                $\mu^i_{p_{opt}}
                                                \leftarrow
                                                \mu^i_{p_{opt}} + 1; $
                                                \STATE $\mu_{p_{opt}} \leftarrow \sum_{i \in I_{p_{opt}}} \mu^i_{p_{opt}}$
					\ENDFOR
					\STATE$A \leftarrow A + 1$
          \STATE \lIf{$\textrm{ICN}$} 
							{
								 $\rho_i \leftarrow \{\rho_i, s\}$
							} \label{alg:cache}
				\ELSE
					\STATE$B \leftarrow B + 1$ \label{alg:block1}
				\ENDIF
			\ELSE
				\STATE$B \leftarrow B + 1$ \label{alg:block2}
			\ENDIF
		\ENDFOR
	\ENDFOR
\end{algorithmic}
\end{algorithm}

A satisfied subscriber for $i$, $s \in \sigma_i'$, acts as a
publisher to the subsequent set of subscribers, which is added
to the existing set of publishers of $i$, $\rho_i$. In this case, a
publisher can either be a caching/replication point or merely a
relay-forwarder. When the publisher is merely a relay-forwarder, a relay-distance cost ($\zeta_p^i$ in line~\ref{alg:dim}) is associated with $i$, which reflects the distance between the relay and the original light-publisher of $i$.
Alternatively, when the publisher is an originating or caching source of the light-publication, the relay-distance cost is set to $0$.
 It should be noted that, the IP model is a special case of the ICN model,
 whereby each request for a light-path represents a unique
 light-publication that is advertised by a single light-publisher and
 ``subscribed to" by a single light-subscriber. Therefore, in the IP
 model, it is only the PA cost function of~(\ref{lcost}) that
 determines the selected route of each light-publication. Furthermore,
 when $D$ of the IP model is transformed into the ICN demand matrix
 $\Delta$, such that each $n$ of $d \in D$ is represented by a set of
 unique light-publications requested by $s$, employing MaxDeg will
 also maximize the number of incoming light-paths to $s$. 

 The complexity of the algorithm is determined by the complexity of
 \emph{Dijkstra's} shortest path algorithm in the main loop,
 iterating over the set of light-publishers $\rho_i \subseteq
 P$.
 Therefore, the lower bound complexity
 (i.e. when $|\rho_i|=1, \ \forall i \in I$) is the same as that of a
 single-source shortest path algorithm,
 $O(|D|(|E| + |V|\log|V|))$~\cite{dij:com}; whereas the upper bound is
 $O(M|S||P|(|E| + |V|\log|V|))$, when
 $|\rho_i| = |P|, \ \forall i \in I $.
 
\section{Evaluation}\label{sec:eva}
Here we evaluate the application of the proposed model on the network performance, reflected by three performance factors: \emph{Blocking Rate}, \emph{Resource Utilization} and \emph{Connectedness}. This has been illustrated by comparing the general utilization improvement gained by ICN, compared to IP. It is further expressed by the improvement gain of MaxDeg over traditional algorithms such as LCP and SFF. For comparing with LCP algorithms, we use the cost function of \cite{ozd:rwa} in MaxDeg algorithm, as opposed to the proposed PA function of~(\ref{lcost}).
For completeness, the SFF algorithm is presented in
Algorithm~\ref{alg:icn_sff} without description, see~\cite{dij:rwa}
for more details. Noticeably, as with MaxDeg (shown in
Algorithm~\ref{alg:icn_maxdeg}), there are differences in
lines~\ref{alg:sffforallrho} and \ref{alg:sffcache} depending upon
whether IP or ICN models are used, as in the IP case there is no support for
\emph{anycast}.
\begin{algorithm}[t]
\caption{$ICN:SFF(G(V,E,W), \Delta)$}
\label{alg:icn_sff}
\begin{algorithmic}[1]
\STATE $A \leftarrow 0; \ B \leftarrow 0;\ F \leftarrow \{1 \mid \forall e \in E \}$
	\FORALL {$\delta_i \in \Delta$}
		\FORALL{$ s \in \sigma_i$}
			\STATE $p_{opt} \leftarrow NA; \ \psi_{opt} \leftarrow \emptyset; \ good \leftarrow \infty;\ best \leftarrow \infty$
            \FORALL[note for IP $|\rho_i|=1$]{$p \in \rho_i$} \label{alg:sffforallrho}
				\STATE$\psi \leftarrow Dijkstra(G,p,s,F)$
				\STATE$good \leftarrow \sum_{e \in \psi} F(e)$
				\IF{$good < best$}
					\STATE$ best \leftarrow good, \ \psi_{opt}\leftarrow \psi, \ p_{opt} \leftarrow p$
				\ENDIF
			\ENDFOR
			\IF {$best \neq \infty$}
				\STATE $\eta \leftarrow FF(\psi_{opt}, U)$
				\IF{$\eta \neq \emptyset $}
					\FORALL {$[e,w] \in \eta$}
						\STATE $U_{e,w} \leftarrow 1$
						\IF{$U_e == \{1 \mid \forall w \in W \}$}
							\STATE$F(e) \leftarrow \infty$
						\ENDIF
					\ENDFOR
					\STATE$A \leftarrow A + 1$
          \STATE \lIf{$\textrm{ICN}$} 
					{
					 $\rho_i \leftarrow \{\rho_i, s\}$
					} \label{alg:sffcache}
				\ELSE
					\STATE$B \leftarrow B+ 1$
				\ENDIF
			\ELSE
				\STATE$B \leftarrow B + 1$
			\ENDIF
		\ENDFOR
	\ENDFOR
\end{algorithmic}
\end{algorithm}

To quantify the benefits of \emph{anycast} and replication features, there is a need to define the number of light-publishers and light-subscribers per light-publication. A frequency distribution model can be utilized to define the size of each set according to the popularity of the light-publication. Determining the best frequency model to fit the popularity of light-publications is a part of our future work. Nevertheless, in the next section we present an example, empirical, model that has proven to apply to a wide range of phenomena. We further justify the selection of this model and utilize it to derive the frequency model of the proposed ICN demand set.

\subsection{A Frequency Distribution Model of Light-Publications}\label{sec:zipf}
Here, we propose a distribution model of the number of nodes that publish/``subscribe to" a light-publication. This model helps us to evaluate the benefits of having multiple light-publishers per light-publication. Since $I$ is a popularity based set of groups of information, the frequency of publishing each $i \in I$ is strongly correlated to the rank of $i$. Recall that the popularity rank of $i$ is defined (in Section~\ref{sec:icndefs}) by the common number of advertisements/subscriptions of raw publications grouped within $i$. Existing research has found strong evidence that, in general, such correlation closely follows a power law where the frequency of occurrence of certain information is inversely proportional to its rank. This applies to the frequency of accessing web servers, the shape and size of aggregate traffic flows, network connectivity, and resource allocations \cite{shi2:Zipf, cav:rwa, nai:rwa,car:ICnet, shi:Zipf, kot:Zipf,bre:Zipf}. 
Particularly, Zipf's law defines the size of the frequency model in terms of the number of occurrences. Here, we utilize Zipf's law to define two correlated frequency models. The first model defines the frequency of advertisement, whereas the second model defines the frequency of subscription. The two models are essentially similar, except for the bound on the number of nodes advertising/``subscribing to" the most popular light-publications.
Now, let us define $k$ as the popularity rank of light-publication $i \in I$, where the size of $I$ is $M$. Next, we define the distribution function of both models as a probability mass function of Zipf's law:
\begin{equation}
L(k,M,\epsilon) = \frac{1/k^\epsilon}{\sum_{m =1}^M 1/m^\epsilon} \label{zipf}
\end{equation}
where $\epsilon$ is the exponent of Zipf's law. Then the number of nodes advertising a light-publication $i \in I$ is defined as a scalar function of (\ref{zipf}):
\begin{equation}
L_p(k,M,\epsilon) = X L(k,M,\epsilon)
\end{equation} 
Recall that publication events do not incur an offered load to the network. It is only when subscriptions are made that light-paths are requested and, therefore, load is offered to the network.
Therefore, the number of nodes advertising a particular light-publication is independent from the total number of light-publications in the network. It is, however, governed by the total number of nodes in the network. The model also ensures that the set of light-publishers of each $i \in I$ does not overlap with the corresponding set of light-subscribers. It is only after a light-subscriber receives $i$, from an originating light-publisher, that it can act as a second publisher (i.e. caching/replication point).

Since subscriptions result in offered load to the network and as a multilayer node acts as a single light-subscriber, the highest number of light-subscribers of $i$ is bounded by either: the total number of demands, $M$, or the number of nodes in the network, $X$. This bound ensures that the overall number of subscriptions reflects the designated offered load to the network.
Accordingly, the number of nodes subscribing to $i \in I$ is defined as:
\begin{equation}
L_s(k,M,\epsilon) = Y  L(k,M,\epsilon) , \ Y = \min(M,X)
\end{equation}
These models have been employed to transform the IP demand set to the ICN demand set, formulated earlier in section~\ref{sec:icnmodel}.

\subsection{Evaluation Model}
Our comparisons have been demonstrated analytically using IP demand sets, which have been synthetically generated following the lognormal model of~\cite{nuc:tm} to emulate Internet level traffic demands. Then, the respective ICN demand sets have been generated by employing the transformation model of Section~\ref{sec:icnmodel} and the frequency model of Section~\ref{sec:zipf}, using a standard Zipf's exponent $\epsilon=1$. Whereby, the number of requested light-paths is used to determine the size of the popularity rank set and generate the frequency models of publications and subscriptions. 
The resultant frequency of publications/subscriptions of the top $10$ light-publications, out of $1000$ offered to the network, is shown in Table~\ref{tab:frq}.
\begin{table}[tb]
  \centering
  \caption{Frequency of publications/subscriptions of the top 10 popular light-publications}
  \label{tab:frq}
  \begin{tabular}{l | c c c c c c c c c c}
  	\hline
    $Frequency$/$Rank$ & $1$ & $2$ & $3$ & $4$ & $5$ & $6$ & $7$ & $8$ & $9$ & $10$\\
     \hline 
    	$Publications$ & $13$ & $6$ & $3$ & $4$ & $3$ & $3$ & $2$ & $2$ & $2$ & $2$\\ 
    $Subscription$ & $14$ & $7$ & $5$ & $4$ & $3$ & $3$ & $2$ & $2$ & $2$ & $2$\\ 
    \hline 
  \end{tabular}
\end{table}

The \emph{blocking rate} and \emph{resource consumption} factors have been measured for two scenarios, modeled using the Geant network topology~\cite{kni:zoo} $G(V=37,E=116,W=36)$. 
The first scenario measures the network performance when the number of
wavelengths per vertex is fixed while the ratio of the offered load to
the total network capacity is increasing. The \emph{offered load} is
defined in terms of a ratio that reflects the total number of
requested light-paths relative to the total number of wavelengths in
the network. The number of wavelengths per vertex has been fixed to
$36$, providing a $X(X-1) = 37\times36$ total wavelength capacity,
while the ratio of offered load increases from $\approxeq 0.33$ to
$\approxeq 0.78$ with a step of $\approxeq 0.05$. These bounds have
been chosen to exemplify a carrier network load, ranging from
relatively low to high.
The second scenario considers a fixed ratio of offered load and
studies the effect of increasing the number of wavelengths. The
offered load is fixed such that there is a demand request from each
vertex to every other vertex in the network (i.e. $X(X-1) =
37\times36$), while the number of wavelengths per vertex increases
from $36$ to $84$ in steps of $12$.

The strength of the resultant light-path topology,
\emph{Connectedness}, has been measured for synthetically generated
mesh graphs; whereby the vertex out-degree follows a Weibull
distribution of a standard shape parameter $k=0.42$ and min/max vertex
degree of $(2,8)$. This factor has been measured for a growing network
size (i.e. increasing number of vertices) or expanding network
capacity (i.e. increasing number of wavelengths). The network growth
follows an incremental model; whereby, the number of vertices starts
from a value of $5$ and increases by a factor of two, while the number
of wavelengths per vertex is fixed to $16$. In the second case, the
network size is fixed to $20$ vertices, while the number of
wavelengths increases from $2^3$ to $2^6$, doubling each time. For all scenarios, each point/bar in the presented figures is the average of $100$ experiments of randomized sets of demands and synthetic graphs, with 95\% confidence interval.

\subsection{Evaluation Results}
\subsubsection{Blocking Rate}
\begin{figure}[tb]
  \centering
  \includegraphics [width=\columnwidth, height=16em,keepaspectratio]{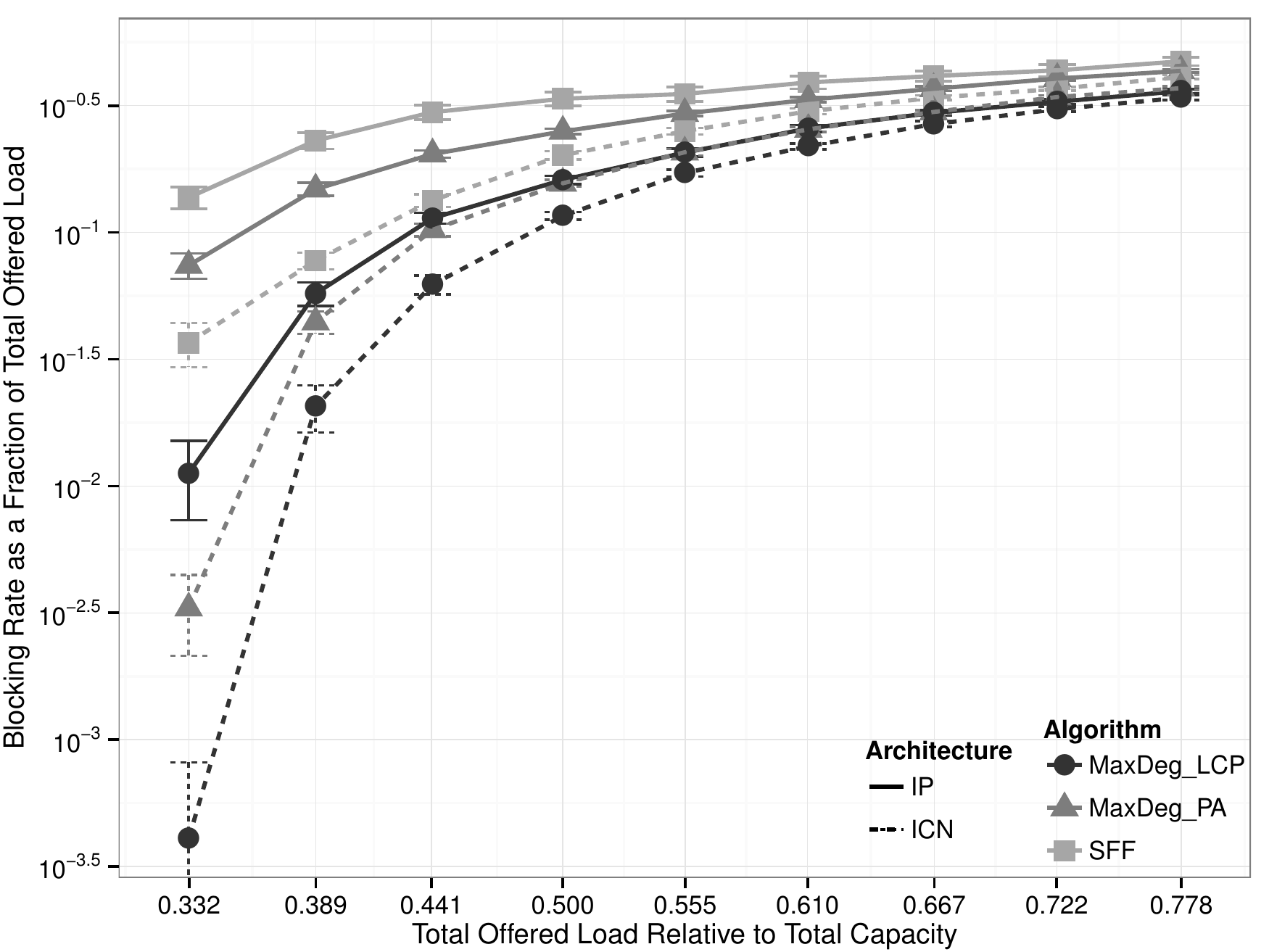}
  \caption{Comparing the average blocking rate of the IP and ICN
    models. The scenario examines the algorithms in the Geant graph
    with $36$ wavelengths and an increasing ratio of offered load.}
  \label{fig:br}
 \end{figure}

\begin{figure}[tb]
    \centering
    \includegraphics [width=\columnwidth, height=16em,keepaspectratio]{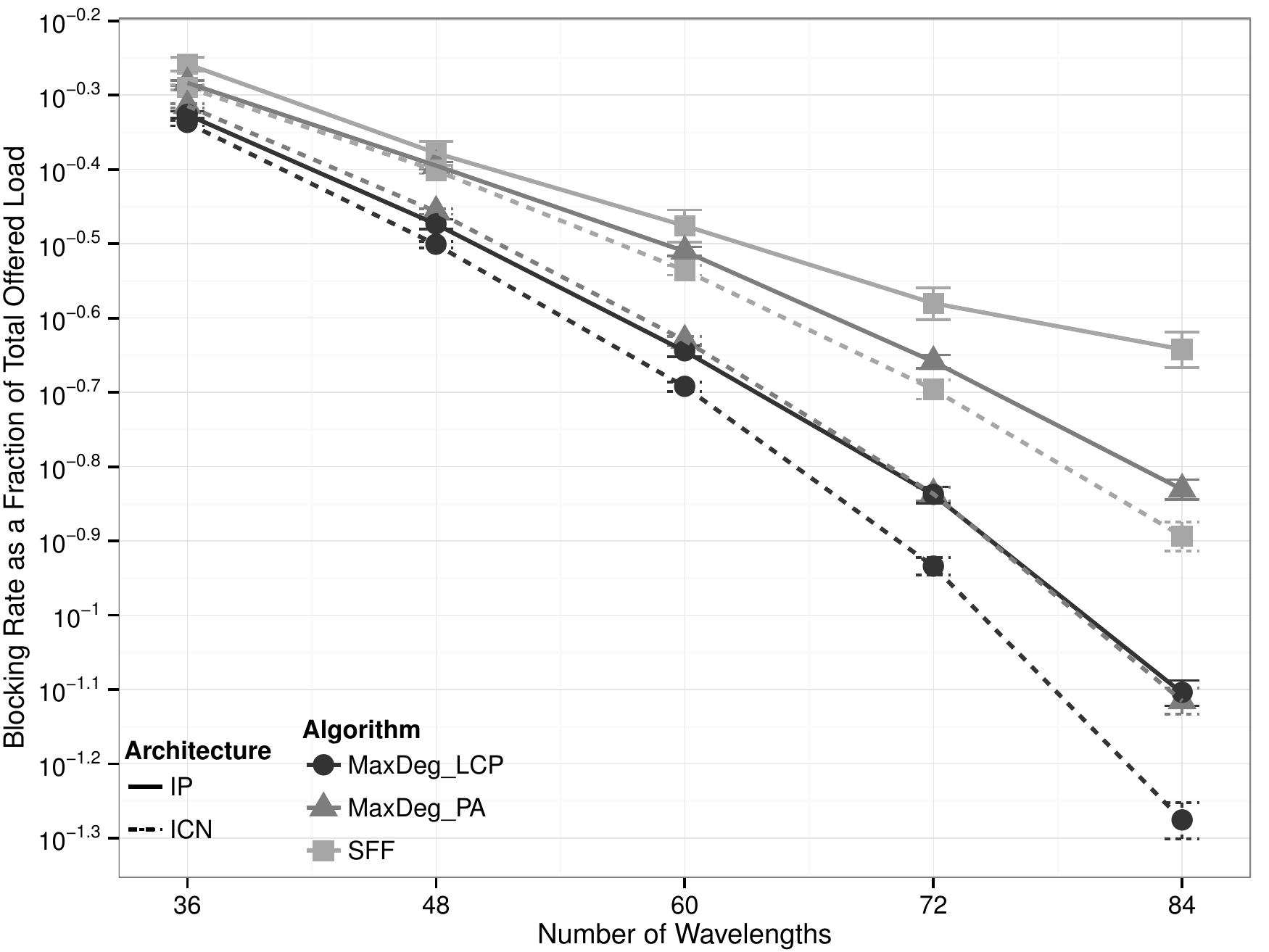}
    \caption{Comparing the average blocking rate of the IP and ICN
    models. The scenario examines the algorithms in the Geant graph
    with $X(X-1)$ demands and an increasing number of wavelengths.}
    \label{fig:ipwbr}
   \end{figure}

This factor reflects the number of rejected light-path requests, measured as a fraction of the ratio of offered load.
The results of the first scenario, presented in \figurename~\ref{fig:br}, indicate that on average when the network is lightly loaded the blocking rate can be decreased by one to two orders of magnitude in the ICN model compared to IP. This has been observed in general, even for the SFF algorithm, which performs considerably better than its IP counterpart. 
The improvement indicates the significant impact of retaining multiple light-publishers per light-publication, either as originating sources or as replication points. The results also show that the blocking rate achieved by MaxDeg\_PA is lower than that of SFF and higher than that of MaxDeg\_LCP. The lower blocking rate comes as a result of the enhanced load balancing of the proposed PA function. However, it is worse than MaxDeg\_LCP due to the concentration effect cause by the power awareness.
 Similarly, when the number of demands is fixed, the blocking rate generally decreases with the increase in the wavelength capacity, as exhibited in \figurename~\ref{fig:ipwbr}. The reduction in the blocking rate is similar to that shown in~\figurename~\ref{fig:br}, albeit with less variation.
 
 \subsubsection{Resource Utilization}
\begin{figure}[tb]
   \centering
   \includegraphics [width=\columnwidth, height=16em,keepaspectratio]{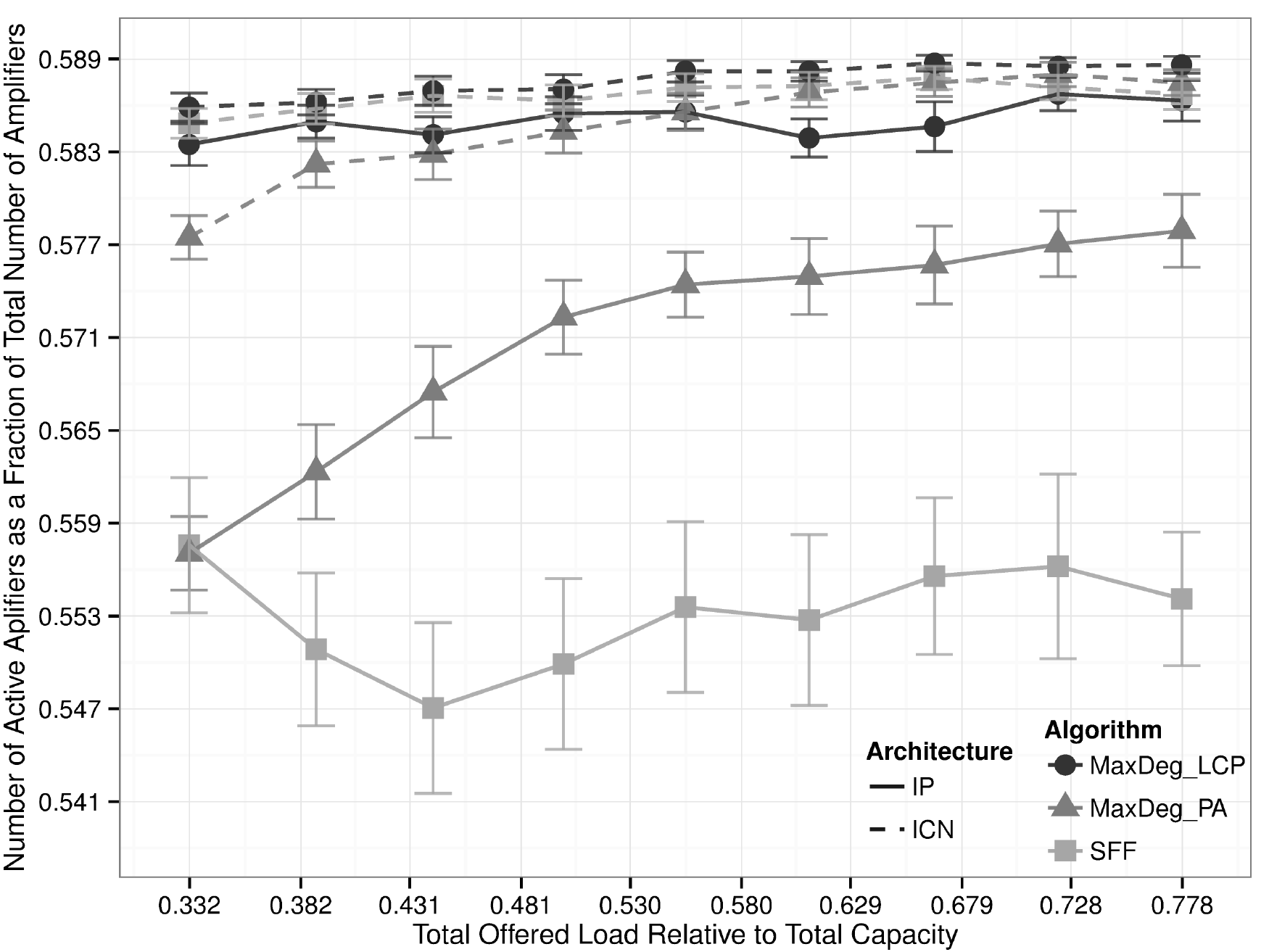}
   \caption{Comparing power consumption in terms of the average ratio
     of active-to-total amplifiers in the IP and ICN models. The
     scenario examines the algorithms in the Geant graph with
     $36$ wavelengths and an increasing ratio of offered load.}
   \label{fig:iplu}
  \end{figure}

 This metric reflects the network cost in terms of power consumption,
 expressed by the number of operating amplifiers, which is controlled
 by the number of utilized links in the network.
 \figurename~\ref{fig:iplu} measures the fraction of amplifiers that
 have been utilized and it shows that all the ICN algorithms have
 similar power consumption to MaxDeg\_LCP in IP. Although, the results seem to
 show that ICN performs worse with regard to power consumption
 compared to SFF and MaxDeg\_PA in IP, this is in the context of the
 ICN model admitting considerably more traffic, as shown in
 \figurename~\ref{fig:br}. The ICN performance is influenced by the
 publisher exhaustion component of the light-publisher cost function, which
 tends to distribute the load of a highly popular light-publication
 over multiple publishers. Consequently, it breaks down the
 corresponding set of light-subscribers into subsets; whereby, a smaller
 delivery tree, which may consists of new links, is established from a
 single light-publisher to each subset. Those links activated by such
 trees are further utilized to deliver less popular
 light-publications, instead of activating new links. This effect is
 further magnified with the use of caching/replication features, which
 expand the set of candidate light-publishers.
 
 \figurename~\ref{fig:iplu} also indicates that the choice of algorithm in the ICN model can be solely based on the blocking rate, since the power variation between the different algorithms is negligible. Although, MaxDeg\_PA shows a noticeable resource saving compared to MaxDeg\_LCP, it is a small gain compared to the reduction in the blocking rate achieved by MaxDeg\_LCP.
In contrast, in the IP model, the ratio of amplifiers lit by MaxDeg\_PA is higher than that of SFF and lower than that of MaxDeg\_LCP. To this extent, the benefit of MaxDeg\_PA in better addressing the discussed trade-off is illustrated in improving the resource savings compared to MaxDeg\_LCP, while maintaining a lower blocking rate than that of SFF.

\begin{figure}[tb]
   \centering
   \includegraphics [width=\columnwidth, height=16em,keepaspectratio]{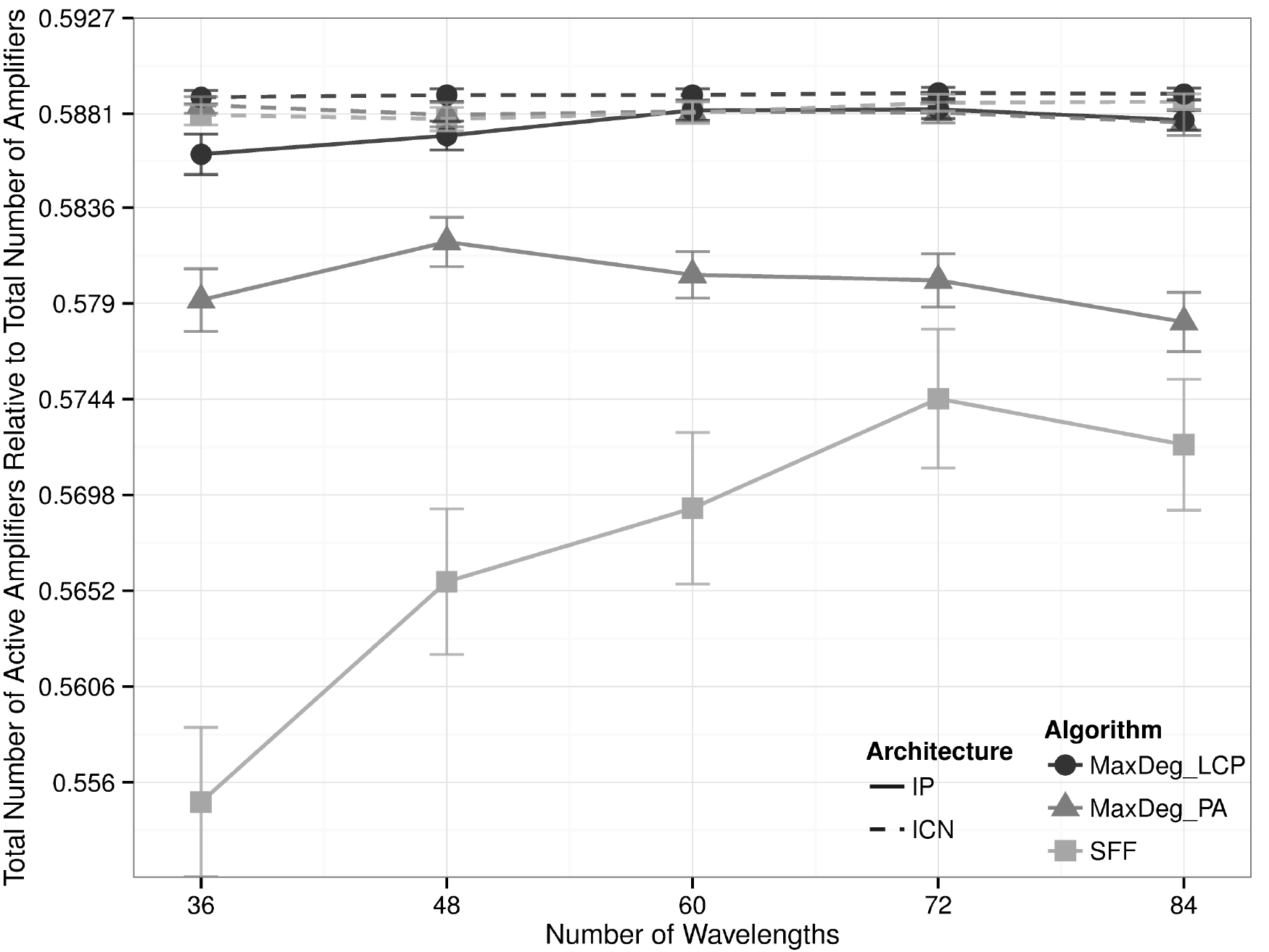}
   \caption{Comparing power consumption in terms of the average ratio
     of active-to-total amplifiers in the IP and ICN models. The
     scenario examines the algorithms in the Geant graph with
     $X(X-1)$ demands and an increasing number of wavelengths.}
   \label{fig:ipwlu}
  \end{figure}

Similar results are exhibited in \figurename~\ref{fig:ipwlu}, for the
ICN model, when increasing the number of wavelengths. However, for the
IP model, the ratio of activated amplifiers by MaxDeg\_PA decreases as
the number of wavelengths increases. This is caused by the
concentration effect of the power awareness term in the proposed
function of~(\ref{lcost}), which benefits from the increased capacity
in routing the admitted demands over fewer links. Whereas, the ratio
activated by MaxDeg\_LCP remains fixed when the number of wavelengths increases. This is caused by the fact that LCP already utilizes as many free edges as it needs, which maximizes the number of operating amplifiers. Therefore, an increase in the number of wavelengths leads to a better network performance, in terms of blocking rate. But, it does not affect the number of utilized links, and therefore the number of amplifiers. 

\subsubsection{Logical Connectedness}
We define \emph{logical connectedness} as the ratio between: the minimum number of light-paths that need to be removed, to disconnect each FW from the other; to the minimum number of fiber links that need to be removed to disconnect each OXC from the other.
This factor provides insight into the strength and density of the logical topology offered to the packet layer, as well as the fairness in satisfying the demands of different nodes. The factor has been measured by the average \emph{connectedness} of the logical topology relative to that of the physical topology, for two ratios of offered load $(0.5,1)$. In this context, a logical edge is established between a pair of packet forwarders if there is at least one light-path that connects them. Accordingly, the capacity of a logical edge is defined by the number of light-paths connecting its end-forwarders (i.e. the flow size).

\paragraph{Expanding Wavelength Capacity}\label{sec:73}

\begin{figure}[tb]
  \centering
  \includegraphics [width=\columnwidth, height=16em,keepaspectratio]{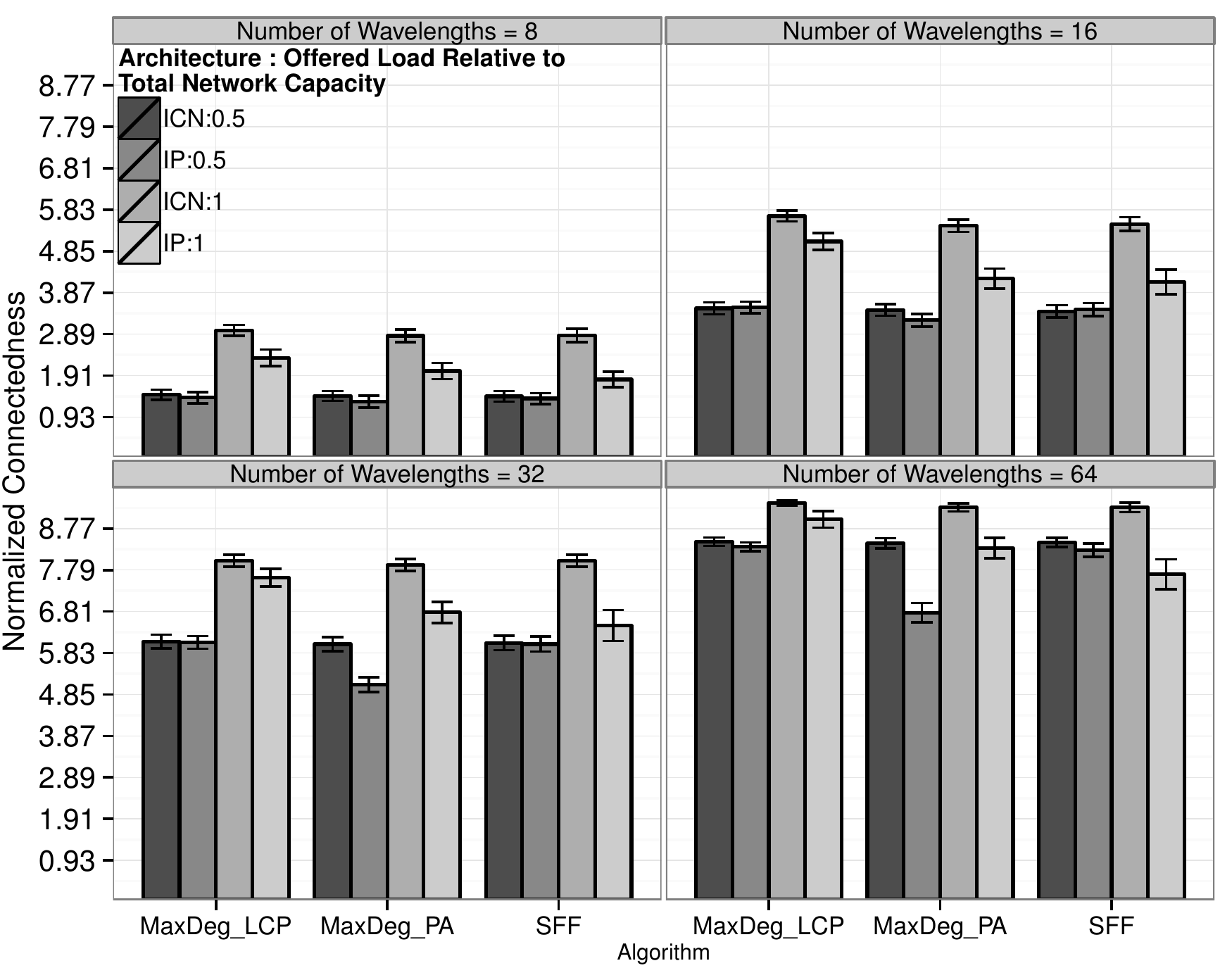}
   \caption{A comparison between the IP and ICN models in terms of light-path topological strength showing the light-path connectedness of a $20$ node synthetic graph with expanding capacity when the offered load ratios are $\{0.5, 1\}$.}
  \label{fig:ipls}
 \end{figure}

\figurename~\ref{fig:ipls} exhibits the average connectedness of the IP and ICN models for fixed network size and expanding wavelength capacity. The results indicate that, in general, the ICN model improves the logical connectedness by $\approxeq 30\%$ for all the examined capacities. Moreover, the variation among the different algorithms in the ICN model is very small, compared to that of the IP model. In the IP model, when the ratio of offered load is 1, the achieved connectedness by MaxDeg\_PA is as good or better than that of SFF. However, when the ratio of offered load is 0.5, the connectedness of MaxDeg\_PA becomes worse than that of SFF as the wavelength capacity increases. This is due to the dominating role of power-awareness in the IP model, which results in fewer logical edges with higher capacity per edge (i.e. more light-paths per edge).

\paragraph{Growing Network Size}

\begin{figure}[tb]
  \centering
  \includegraphics [width=\columnwidth, height=16em,keepaspectratio]{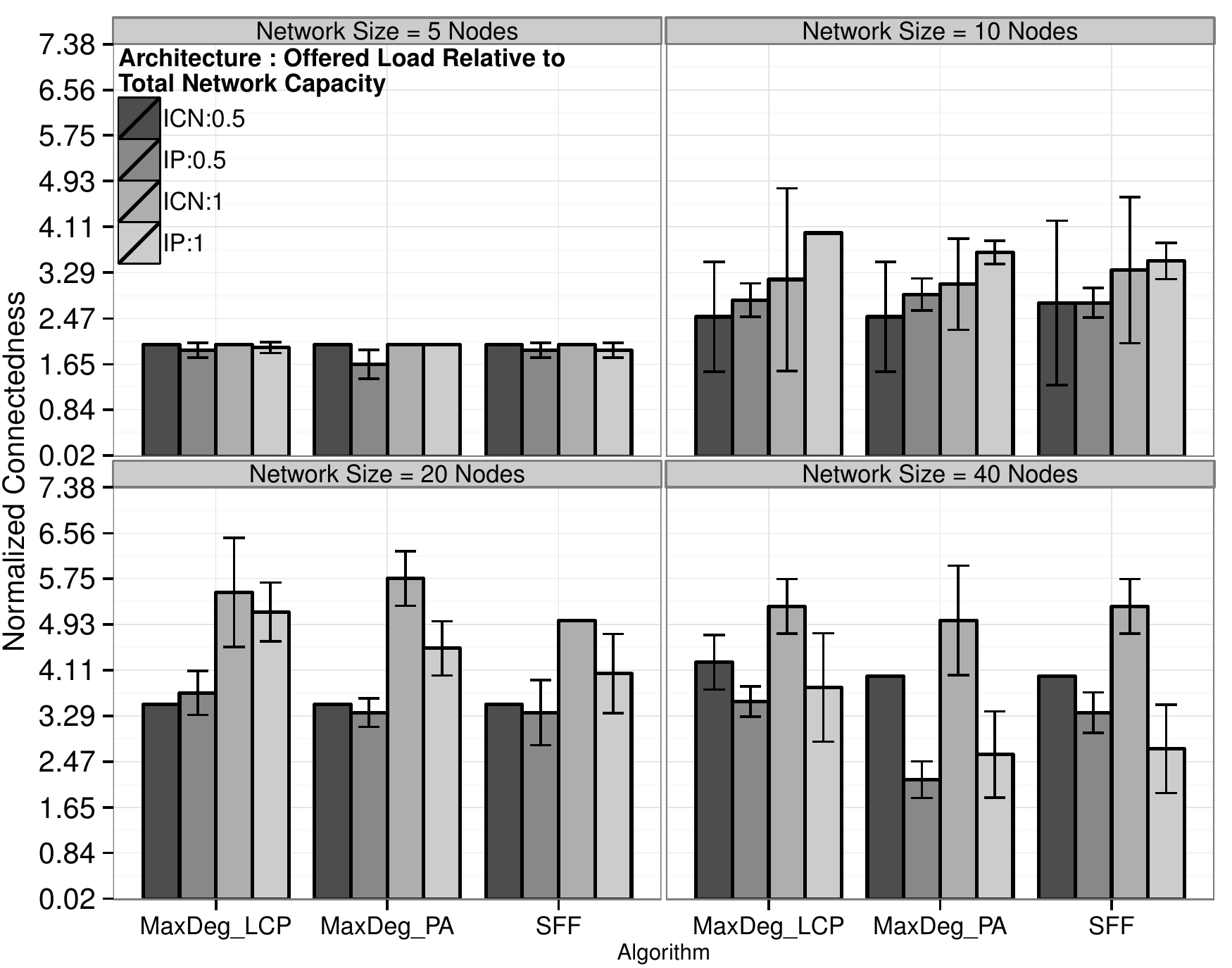}
  \caption{A comparison between the IP and ICN models in terms of light-path topological strength showing the light-path connectedness of a growing size graph with fixed wavelength capacity of $16$ colors per vertex when the offered load ratios are $\{0.5, 1\}$.}
  \label{fig:ipwls}
 \end{figure}

\figurename~\ref{fig:ipwls} shows the average connectedness of the IP and ICN models for a growing size graph with fixed wavelength capacity ($Z=16$). The results indicate that for a small network ($5$ nodes), the connectedness is similar for all the examined algorithms. This is due to the excessive wavelength capacity compared to the graph size, which exceeds the capacity needed to achieve full mesh connectivity. Therefore, increasing the number of admitted light-paths beyond $|V| (|V| -1)/2$ will not increase the logical connectedness; although, it will increase the capacity per logical edge.
In contrast, for medium sized networks ($40$ nodes), the connectedness
generally drops, compared to that of smaller networks ($20$
nodes). This is because the min/max vertex degree does not increase with
the graph size. Therefore, increasing the number of vertices is met by a
proportional decrease in the physical reachability of each vertex. When
the wavelength capacity is fixed, this also results in wavelength
scarcity, which decreases the likelihood of finding a route that satisfies the wavelength continuity constraint.
 
 Additionally, when analyzing the performance of our model using
 synthetically generated sparse graphs (omitted here due to shortage
 of space): it has been observed that, for the IP model, substantial
 power savings can be achieved for slightly poorer blocking rate
 compared to SFF. However, for the ICN model considerable power
 savings can be achieved for similar blocking rate as that of
 LCP. This is because our algorithm benefits from the consistent
 load-balancing advantage of LCP; thereby, it avoids bottleneck
 scenarios created by SFF, which only aims to minimize provisioned
 resources. At the same time, our algorithm achieves such
 load-balancing with fine tuning of path choices towards already lit
 paths, which can provide considerable power saving compared to LCP.

 The results show a clear advantage of ICN/WDM over IP/WDM. As
 the algorithm used to generate the results is identical for both, we can state that the performance advantages of
 ICN/WDM are due to the ability of ICN to support multiple
 publishers per light-publication. This is enabled by the dual features of ICN:
 naming and decoupling of content identity from location. These
 features are not provided by the IP architecture. Furthermore, ICN shows greater connectivity, thus
 providing higher resilience opportunities.
 
\section{Conclusion}\label{sec:con}
Efficient multilayer resource utilization is a success factor for future Internet architectures. In this paper, we introduced a novel multilayer architecture in the form of an information-centric WDM network (ICN/WDM). The proposed architecture utilizes an ICN pub/sub communication model to facilitate traffic aggregation based on information semantics, rather than end point location. The work is demonstrated using the PURSUIT architecture although the general principle applies to other ICN architectures. The work makes use of common ICN features such as native support of \emph{anycast} as well as caching/replication functions, which leads to enhanced, power efficient, network utilization. Furthermore, a novel power-aware (PA) algorithm has been proposed to address the trade-off between power consumption and performance. Evaluation results have shown the general superiority of the ICN/WDM architecture, compared to that of IP/WDM. Furthermore, the results illustrate the advantages of the proposed algorithm in improving the utilization of the IP/WDM architecture.
Notably, the ICN/WDM architecture can achieve considerably better performance using an LCP-based algorithm, for similar power consumption to that of a PA algorithm.

\bibliographystyle{IEEEtran}

\begin{thebibliography}{45}
\providecommand{\url}[1]{#1}
\csname url@samestyle\endcsname
\providecommand{\newblock}{\relax}
\providecommand{\bibinfo}[2]{#2}
\providecommand{\BIBentrySTDinterwordspacing}{\spaceskip=0pt\relax}
\providecommand{\BIBentryALTinterwordstretchfactor}{4}
\providecommand{\BIBentryALTinterwordspacing}{\spaceskip=\fontdimen2\font plus
\BIBentryALTinterwordstretchfactor\fontdimen3\font minus
  \fontdimen4\font\relax}
\providecommand{\BIBforeignlanguage}[2]{{%
\expandafter\ifx\csname l@#1\endcsname\relax
\typeout{** WARNING: IEEEtran.bst: No hyphenation pattern has been}%
\typeout{** loaded for the language `#1'. Using the pattern for}%
\typeout{** the default language instead.}%
\else
\language=\csname l@#1\endcsname
\fi
#2}}
\providecommand{\BIBdecl}{\relax}
\BIBdecl

\bibitem{tro:ICnet}
D.~Trossen, M.~S\"{a}rel\"{a}, and K.~Sollins, ``Arguments for an
 information-centric internetworking architecture,'' \emph{ACM SIGCOMM
 Computer Communications Review}, vol.~40, pp. 26--33, Apr 2010.


\bibitem{jac:ICnetCACM}
V.~Jacobson, D.~K. Smetters, J.~D. Thornton, M.~Plass, N.~Briggs, and
 R.~Braynard, ``Networking named content,'' \emph{Commun. ACM}, vol.~55,
 no.~1, pp. 117--124, Jan 2012.

\bibitem{kop:ICnet}
T.~Koponen, M.~Chawla, B.~Chun, A.~Ermolinskiy, K.~H. Kim, S.~Shenker, and
 I.~Stoica, ``A data-oriented (and beyond) network architecture,'' \emph{ACM
 SIGCOMM Computer Communications Review}, vol.~37, no.~4, pp. 181--192, Aug
 2007.

\bibitem{xyl:ICnet}
G.~Xylomenos, C.~Ververidis, V.~Siris, N.~Fotiou, C.~Tsilopoulos, X.~Vasilakos,
 K.~Katsaros, and G.~Polyzos, ``A survey of information-centric networking
 research,'' \emph{Communications Surveys Tutorials, IEEE}, vol.~16, no.~2,
 pp. 1024--1049, Feb 2014.

\bibitem{lab:trf}
C.~Labovitz, S.~Iekel-Johnson, D.~McPherson, J.~Oberheide, and F.~Jahanian,
 ``Internet inter-domain traffic,'' \emph{ACM SIGCOMM Computer Communications
 Review}, vol.~41, no.~4, Aug 2011.

\bibitem{pop:ICnet}
L.~Popa, A.~Ghodsi, and I.~Stoica, ``Http as the narrow waist of the future
 internet,'' in \emph{Proc. of 9th ACM SIGCOMM Workshop on Hot Topics in
 Networks, HOTNETS'10}, ser. Hotnets-IX.\hskip 1em plus 0.5em minus
 0.4em\relax ACM, 2010, pp. 1--6.

\bibitem{moc:trf}
K.~Mochalski and H.~Schulze, ``{Ipoque Internet Study 2008/2009},''
 \emph{Africa}, Jan 2009.

\bibitem{kih:trf}
M.~Kihl, P.~Odling, C.~Lagerstedt, and A.~Aurelius, ``Traffic analysis and
 characterization of internet user behavior,'' in \emph{Proc. of Int. Congress
 on Ultra Modern Telecommunications and Control Systems and Workshops,
 ICUMT'10}, Oct 2010, pp. 224--231.

\bibitem{tro12:ICnet}
D.~Trossen, J.~Riihij{\"a}rvi, P.~Nikander, P.~Jokela, J.~Kj{\"a}llman, and
 J.~Rajahalme, ``Designing, implementing and evaluating a new internetworking
 architecture,'' \emph{Computer Commun.}, vol.~35, no.~17, pp.
 2069--2081, 2012.



\bibitem{pet:ICnet}
P.~Jokela, A.~Zahemszky, C.~E. Rothenberg, S.~Arianfar, and P.~Nikander,
 ``Lipsin: line speed publish/subscribe inter-networking,'' \emph{ACM SIGCOMM
 Comp. Commun. Rev.}, vol.~39, pp. 195--206, 2009.

\bibitem{wia:alloptic}
P.~Wiatr, P.~Monti, and L.~Wosinska, ``Power savings versus network performance
 in dynamically provisioned wdm networks,'' \emph{IEEE Commun. Mag.}, vol.~50, no.~5, pp. 48--55, May 2012.

\bibitem{coi:alloptic}
A.~Coiro, M.~Listanti, and A.~Valenti, ``Dynamic Power-Aware Routing and Wavelength Assignment for Green WDM Optical Networks,'' in
 \emph{IEEE Int. Conf. Commun. (ICC), 2011}, Jun 2011,
 pp. 1--6.

\bibitem{raj:alloptic}
R.~Ramaswami and K.~N. Sivarajan, \emph{Optical Networks: A Practical
 Perspective}.\hskip 1em plus 0.5em minus 0.4em\relax Kaufmann, Morgan, 2009.

\bibitem{zan:alloptic}
H.~Zang, J.~P. Jue, and B.~Mukherjee, ``A review of routing and wavelength
 assignment approaches for wavelength-routed optical wdm networks,''
 \emph{Optical Networks Magazine}, vol.~1, pp. 47--60, 2000.

\bibitem{and:rwa}
M.~Andrews and L.~Zhang, ``Complexity of wavelength assignment in optical
 network optimization,'' \emph{IEEE/ACM Transactions on Networking}, vol.~17,
 no.~2, pp. 646--657, Apr 2009.

\bibitem{ram:rwa}
R.~Ramaswami and K.~N. Sivarajan, ``Routing and wavelength assignment in
 all-optical networks,'' \emph{IEEE/ACM Transactions on Networking}, vol.~3,
 no.~5, pp. 489--500, Oct 1995.

\bibitem{ban:rwa}
D.~Banerjee and B.~Mukherjee, ``Wavelength-routed optical networks: linear
 formulation, resource budgeting tradeoffs, and a reconfiguration study,''
 \emph{IEEE/ACM Transactions on Networking}, vol.~8, no.~5, pp. 598--607, Oct
 2000.

\bibitem{dij:rwa}
E.~W. Dijkstra, ``A note on two problems in connexion with graphs,''
 \emph{Numerical Mathematics}, vol.~1, no.~1, pp. 269--271, 1959.
 
 \bibitem{dij:com}
 M. Barbehenn, ``A note on the complexity of Dijkstra's algorithm for graphs with weighted vertices," 
 \emph{IEEE Transactions on Computers}, vol.~47, no.~2, pp. 263, Feb 1998

\bibitem{ozd:rwa}
A.~E. Ozdaglar and D.~P. Bertsekas, ``Routing and wavelength assignment in
 optical networks,'' \emph{IEEE/ACM Transactions on Networking}, vol.~11,
 no.~2, pp. 259--272, Apr 2003.

\bibitem{cha:rwa}
K.~Chan and T.~P. Yum, ``Analysis of least congested path routing in wdm
 lightwave networks,'' in \emph{INFOCOM '94. Networking for Global
 Communications., 13th Proceedings IEEE}, Jun 1994, pp. 962--969 vol.2.

\bibitem{chu:rwa}
X.~Chu and B.~Li, ``Dynamic routing and wavelength assignment in the presence
 of wavelength conversion for all-optical networks,'' \emph{IEEE/ACM
 Transactions on Networking}, vol.~13, no.~3, pp. 704--715, Jun 2005.

\bibitem{yon:rwa}
Y.~Wu, L.~Chiaraviglio, M.~Mellia, and F.~Neri, ``Power-aware routing and
 wavelength assignment in optical networks,'' in \emph{Optical Commun.,
 2009. ECOC '09. 35th European Conference on}, Sep 2009, pp. 1--2.

\bibitem{coi2:rwa}
A.~Coiro, M.~Listanti, A.~Valenti, and F.~Matera, ``Reducing power consumption
 in wavelength routed networks by selective switch off of optical links,''
 \emph{Selected Topics in Quantum Electronics, IEEE Journal of}, vol.~17,
 no.~2, pp. 428--436, Mar 2011.


\bibitem{mjr:te}
M.~J. Reed, ``Traffic engineering for information-centric networks,'' in
 \emph{ICC}, 2012, pp. 2660--2665.

\bibitem{nad:ICnet}
M.~F. Al-Naday, M.~J. Reed, D.~Trossen, and K.~Yang, ``Information resilience:
 source recovery in an information-centric network,'' \emph{IEEE Network},
 vol.~28, no.~3, pp. 36--42, May 2014.

\bibitem{sou2:zipf}
V.~Sourlas, L.~Gkatzikis, P.~Flegkas, and L.~Tassiulas, ``Autonomic cache
 management in information-centric networks,'' in \emph{Network Operations and
 Management Symposium (NOMS), IEEE}, Apr 2012, pp. 121--129.

\bibitem{Zha:te}
Q.~Zhao, Z.~Ge, J.~Wang, and J.~Xu, ``Robust traffic matrix estimation with
 imperfect information: making use of multiple data sources,''
 \emph{SIGMETRICS Perform. Eval. Rev.}, vol.~34, no.~1, pp. 133--144, Jun
 2006.

\bibitem{kar:te}
T.~Karagiannis, M.~Molle, M.~Faloutsos, and A.~Broido, ``A nonstationary
 poisson view of internet traffic,'' in \emph{IEEE Conference on Computer
 Communications (INFOCOM) 2004}, Mar 2004, pp. 1558--1569.

\bibitem{nuc:tm}
A.~Nucci, A.~Sridharan, and N.~Taft, ``The problem of synthetically generating
 ip traffic matrices: initial recommendations,'' \emph{ACM SIGCOMM Computer
 Communications Review}, vol.~35, no.~3, pp. 19--32, Jul 2005.

\bibitem{ste:rwa}
T.~E. Stern and K.~Bala, \emph{Multiwavelength Optical Networks: A Layered
 Approach}.\hskip 1em plus 0.5em minus 0.4em\relax Boston, MA, USA:
 Addison-Wesley Longman Publishing Co., Inc., 1999.

\bibitem{ahu:netflow}
R.~K. Ahuja, T.~L. Magnanti, and J.~B. Orlin, \emph{Network Flows: Theory,
 Algorithms, and Applications}.\hskip 1em plus 0.5em minus 0.4em\relax Upper
 Saddle River, NJ, USA: Prentice-Hall, Inc., 1993.

\bibitem{cha:omc}
N.~Charbonneau and V.~M. Vokkarane, ``Static routing and wavelength assignment
 for multicast advance reservation in all-optical wavelength-routed wdm
 networks,'' \emph{IEEE/ACM Transactions on Networking}, vol.~20, no.~1, pp.
 1--14, Feb 2012.

\bibitem{tap:bf}
J.~Tapolcai, A.~Gulyás, Z.~Heszberger, J.~Biro, P.~Babarczi, and D.~Trossen,
 ``Stateless multi-stage dissemination of information: Source routing
 revisited,'' in \emph{Global Communications Conference (GLOBECOM), 2012
 IEEE}, Dec 2012, pp. 2797--2802.


\bibitem{shi2:Zipf}
L.~Shi, Z.~Gu, L.~Wei, and Y.~Shi, ``Quantitative analysis of zipf's law on web
 cache,'' in \emph{Proceedings of the Third International Conference on
 Parallel and Distributed Processing and Applications}, ser. ISPA'05.\hskip
 1em plus 0.5em minus 0.4em\relax Springer-Verlag, 2005, pp. 845--852.

\bibitem{cav:rwa}
D.~Cavendish, A.~Kolarov, and B.~Sengupta, ``Is it a good idea to design wdm
 networks to minimize the number of wavelengths used?'' in
 \emph{Communications, 2004 IEEE International Conference on}, vol.~4, Jun
 2004, pp. 2097--2101 Vol.4.

\bibitem{nai:rwa}
S.~Naiksatam, S.~Figueira, S.~A. Chiappari, and N.~Bhatnagar, ``Analyzing the
 advance reservation of lightpaths in lambda-grids,'' in \emph{Cluster
 Computing and the Grid, 2005. CCGrid 2005. IEEE International Symposium on},
 vol.~2, May 2005, pp. 985--992 Vol. 2.

\bibitem{car:ICnet}
G.~Carofiglio, V.~Gehlen, and D.~Perino, ``Experimental evaluation of memory
 management in content-centric networking,'' in \emph{Communications (ICC),
 2011 IEEE International Conference on}, Jun 2011, pp. 1--6.

\bibitem{shi:Zipf}
L.~Shi, Z.~Gu, L.~Wei, and Y.~Shi, ``An applicative study of zipf?s law on web
 cache,'' \emph{International Journal of Information Technology}, vol.~12,
 no.~4, pp. 49--58, 2006.

\bibitem{kot:Zipf}
I.~Kotera, R.~Egawa, H.~Takizawa, and H.~Kobayashi, ``Modeling of cache access
 behavior based on zipf's law,'' in \emph{Proceedings of the 9th Workshop on
 MEmory Performance: DEaling with Applications, Systems and Architecture},
 ser. MEDEA '08.\hskip 1em plus 0.5em minus 0.4em\relax ACM, 2008, pp. 9--15.

\bibitem{bre:Zipf}
L.~Breslau, P.~Cao, L.~Fan, G.~Phillips, and S.~Shenker, ``Web caching and
 zipf-like distributions: evidence and implications,'' in \emph{INFOCOM '99.
 Eighteenth Annual Joint Conference of the IEEE Computer and Communications
 Societies. Proceedings. IEEE}, vol.~1, Mar 1999, pp. 126--134 vol.1.

\bibitem{kni:zoo}
S.~Knight, H.~X. Nguyen, N.~Falkner, R.~Bowden, and M.~Roughan, ``The internet
 topology zoo,'' \emph{IEEE Journal on Selected Areas in Communications},
 vol.~29, pp. 1765--1775, 2011.
 
\bibitem{mud:pow}
T.~Mudge, ``{Power: A First Class Design Constraint for Future
Architectures},'' in \emph{High Performance Computing HiPC 2000}, ser.
Lecture Notes in Computer Science, M.~Valero, V.~Prasanna, and S.~Vajapeyam,
Eds.\hskip 1em plus 0.5em minus 0.4em\relax Springer Berlin Heidelberg, 2000,
vol. 1970, pp. 215--224.

\bibitem{zha:pow}
Y.~Zhang, P.~Chowdhury, M.~Tornatore, and B.~Mukherjee, ``{Energy Efficiency in
 Telecom Optical Networks},'' \emph{IEEE Communications Surveys Tutorials},
 vol.~12, no.~4, pp. 441--458, Nov 2010.
\end{thebibliography}

\begin{IEEEbiography}[{\includegraphics[width=1in,height=1.25in,clip,keepaspectratio]{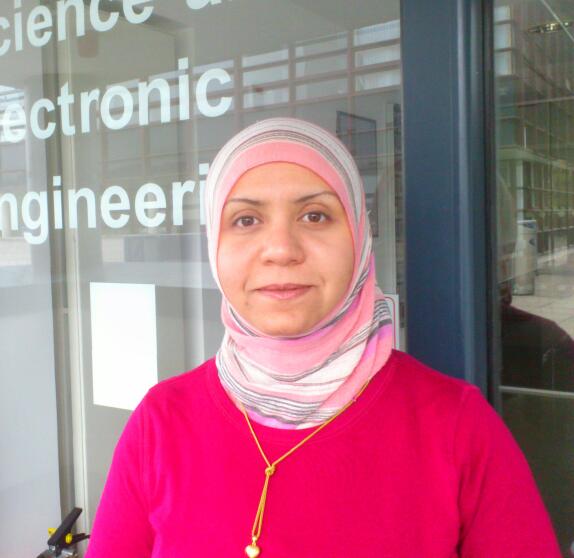}}]{Mays F. AL-Naday}
is a researcher at the School of Computer Science and Electronic Engineering, University of Essex. She received her BSc degree in Communication and Information Engineering from the University of Baghdad. She received her Master's degree in Computer and Information Networks from the University of Essex, after which she has been awarded the University of Essex scholarship for PhD. During her PhD, she joined the Network Convergence Laboratory and worked as a researcher in number of projects, including EU one. Her current research interests involve future Internet architectures, Information-centric Networks, all-optical networks, and multilayer network architectures.
\end{IEEEbiography}
\begin{IEEEbiography}[{\includegraphics[width=1in,height=1.25in,clip,keepaspectratio]{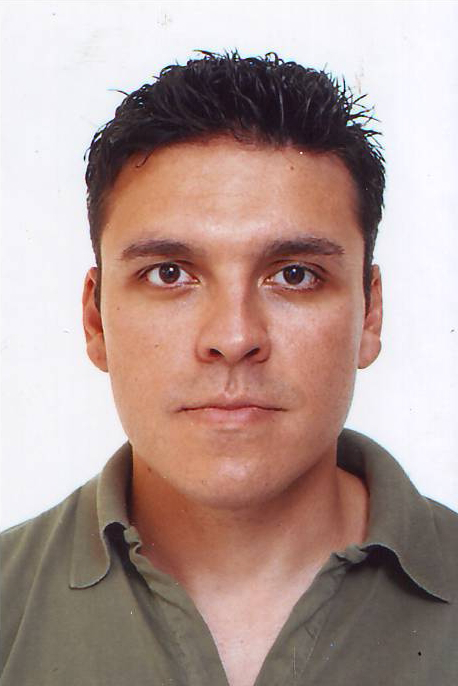}}]{N. Thomos}
	(S'02-M'06-SM'16) received the Diploma and the Ph.D. degrees from the Electrical and Computer Engineering Department of the Aristotle University of Thessaloniki, Thessaloniki, Greece, in 2000 and 2005, respectively. Currently he is Lecturer at the University of Essex, Colchester, United Kingdom. Before that, he was senior researcher with Signal Processing Laboratory at Swiss Federal Institute of Technology (EPFL), Lausanne, Switzerland. His research interests include network coding, multimedia communications, joint source and channel coding and distributed source coding. In 2008, he has been awarded the highly esteemed Ambizione career award from Swiss National Science Foundation targeted to prospective researchers.
\end{IEEEbiography}
\begin{IEEEbiography}[{\includegraphics[width=1in,height=1.25in,clip,keepaspectratio]{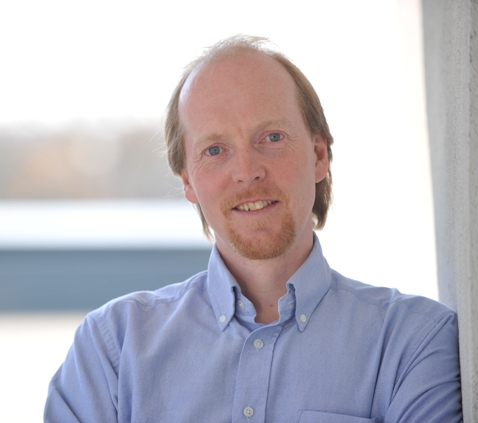}}]{Martin J. Reed} is a Senior Lecturer at the University of Essex. He received his PhD from University of Essex, UK, in
1998. After working as a research officer in the area of multimedia
transmission over IP/ATM networks he was appointed as a lecturer at the
University of Essex in 1998. His research interests include transmission
of media, including ultra-high definition video, over networks, the control
of transport networks, information centric networking and
network security. He has been involved in a number of EPSRC, EU and Industrial projects in these areas and has held a Research Fellowship at BT.
\end{IEEEbiography}
\vfill

\end{document}